\documentclass[twocolumn]{aastex631}

\usepackage[T1]{fontenc}
\usepackage{amsmath}

\graphicspath{{./}{fig/}}

\shorttitle{Magnetic Pressure Saturated Disk with Outflow}
\shortauthors{Huang et al.}

\begin{document}

\title{Black hole accretion with saturated magnetic pressure and disk wind}

\author[0000-0001-8674-2336]{Jiahui Huang}
\affiliation{Department of Engineering Physics, Tsinghua University, Beijing 100084, China}

\author[0000-0001-7584-6236]{Hua Feng}
\email{hfeng@tsinghua.edu.cn}
\affiliation{Department of Astronomy, Tsinghua University, Beijing 100084, China}

\author[0000-0003-3137-1851]{Wei-Min Gu}
\affiliation{Department of Astronomy, Xiamen University, Xiamen, Fujian 361005, China}

\author{Wen-Biao Wu}
\affiliation{Department of Astronomy, Xiamen University, Xiamen, Fujian 361005, China}

\begin{abstract} 
We construct an analytical black hole accretion disk model that incorporates both magnetic pressure and disk wind, which are found to be important from numerical simulations. A saturated magnetic pressure that relates the Alfv\'{e}n velocity with local Keplerian velocity and gas sound speed is assumed in addition to radiation and gas pressures. The mass accretion rate is assumed to have a power-law form in response to mass loss in the wind. We find three sets of self-consistent solutions that are thermally stable and satisfy the model assumptions. At high accretion rates, the disk is geometrically and optically thick, resembling the slim disk solution. At relatively low accretion rates, our model predicts an accretion flow consisting of a geometrically thin and optically thick outer disk (similar to the standard disk), and a geometrically thick and optically thin inner disk (similar to the advection-dominated accretion flow or ADAF). Thus, this is a natural solution for a truncated disk connected with an inner ADAF, which has been proposed to explain some observations. The magnetic pressure plays a more important role than the outflow in shaping the disk structure. The observed disk luminosity tends to saturate around 8 times the Eddington limit, suggesting that supercritical accretion onto black holes can be used for black hole mass estimate, or a standard candle with known black hole masses.
\end{abstract}

\section{Introduction}

Accretion onto black holes can release substantial electromagnetic radiation, manifesting themselves as X-ray binaries (XRBs) or active galactic nuclei (AGNs). Over the past decades, a number of analytical models have been developed to describe the underlying physical processes in the accretion flow. The Shakura-Sunyaev disk (SSD), also known as the standard disk model \citep{Shakura1973}, assumes a balance between the viscous heating and local thermal radiation from an optically thick, geometrically thin accretion disk that is fed at a moderate accretion rate. At high accretion rates, if the advective cooling dominates over radiative cooling, an optically thick, geometrically slim accretion disk can be constructed \citep{Abramowicz1988}. When the accretion rate is extremely low, the disk may be optically thin and radiatively inefficient, forms an advection-dominated accretion flow (ADAF), which is hot and geometrically thick \citep{Narayan1994, Narayan1995}. These models, among many others \citep[e.g.,][]{Abramowicz1978, Narayan2002, Begelman2012}, have been widely used to depict emission from XRBs and AGNs at various emission states \citep{Remillard2006, Yuan2014}.

Most of the above analytical models do not consider disk winds or outflows. However, such outflows are observed to be prevalent in accreting compact objects, e.g., in Galactic XRBs \citep{Lee2002}, ultraluminous X-ray sources \citep[ULXs;][]{Pinto2016}, quasars \citep{Murray1995}, and low-luminosity AGNs \citep{Wang2013}. Magnetohydrodynamic (MHD) simulations have confirmed that the accretion disk can produce significant outflows no matter in what accretion status, including the radiatively inefficient hot accretion flows \citep{stone1999, Ohsuga2011, Yuan2012a, Yuan2012, Yuan2015}, the standard thin disk \citep{Ohsuga2011, Nomura2016, Nomura2020, Huang2023}, or regimes with near or super-critical accretion \citep{Ohsuga2011, Jiang2014, Sadowski2014, Sadowski2015, Kitaki2017, Kitaki2018, Jiang2019, Huang2023}. In order to account for outflows in the analytical model, \citet{Blandford1999} assumed that the accretion rate has a power-law dependence on radius. Based on this assumption, many works have been carried out to study the structure and cooling mechanisms of the accretion disk with outflows \citep[e.g.,][]{Gu2015, Ghoreyshi2020, Wu2022}. Recently, 2D analytical models revealed that the mass accretion rate indeed follows a power-law form \citep{Xie2008, Dotan2011, Gu2012, Cao2015, Kumar2018, Feng2019, Mosallanezhad2021, Zeraatgari2021}.

The magnetic field is another key element that could substantially affect the dynamics and structure of an accretion disk. The magnetorotational instability (MRI) is believed to drive angular momentum transfer in the accretion disk \citep{Balbus1991}. Numerical simulations have shown that the magnetic pressure plays an important role in supporting the disk \citep{Sadowski2016, MoralesTeixeira2018, Jiang2019a, Jiang2019, Lancova2019, Wielgus2022, Huang2023}. However, because of its turbulent nature, some simplifications are needed to add the magnetic fields into the analytical model. \citet{Pariev2003} assumed that the strength of magnetic field varies with radius as a power-law function. \citet{Begelman2007} investigated the structure of a disk supported by saturated magnetic pressure. Recently, more analytical works have been created to discuss the thermal stability of magnetically supported accretion disks \citep{Oda2009, Zheng2011, Yu2015}. 

Recently, with radiative MHD simulations, \citet{Huang2023} found that the saturated magnetic pressure \citep{Begelman2007} is a reasonable approximation, which is the direct motivation of this work. We thus try to construct an analytical model taking into account both disk wind and magnetic pressure, following the assumptions in \citet{Blandford1999} and \citet{Begelman2007} for the two components, respectively, and find the thermal equilibrium solutions following the recipe described in \citet{Wu2022}. The paper is organized as follows. We describe the basic equations in Section \ref{sec:equations}, present the numerical results in Section \ref{sec:results}, and discuss our results in Section \ref{sec:discussion}.

\section{Basic Equations}
\label{sec:equations}

We consider an axisymmetric accretion flow under the pseudo-Newtonian potential \citep{Paczynsky1980},
\begin{equation}
\Phi = - \frac{GM_{\rm BH}}{R-R_{\rm s}} \, ,
\end{equation}
where $G$ is the gravitational constant, $M_{\rm BH}$ is the mass of the central black hole, and $R_{\rm s}\equiv2GM_{\rm BH}/c^2$ is the Schwarzschild radius. The corresponding local Keplerian velocity $V_{\rm K}$ and local Keplerian angular velocity $\Omega_{\rm K}$ at radius $R$ are $V_{\rm K}=\sqrt{\frac{GM_{\rm BH}R}{\left(R-R_{\rm s}\right)^2}}$ and $\Omega_{\rm K}=\sqrt{\frac{GM_{\rm BH}}{R(R - R_{\rm s})^2}}$, respectively.

The basic hydrodynamic equations are used to describe a steady-state axisymmetric accretion flow with disk wind:
\begin{equation}
\label{equ:continuity}
\frac{1}{R} \frac{d}{dR} \left( R \Sigma V_R \right) + \frac{1}{2 \pi R} \frac{d\dot{M}_{\rm w}}{dR} = 0 \, ,
\end{equation}
\begin{equation}
\label{equ:momentum_R}
V_R \frac{dV_R}{dR} + \left( \Omega_{\rm K}^2 - \Omega^2 \right) R + \frac{1}{\rho} \frac{dP}{dR} = 0 \, ,
\end{equation}
\begin{equation}
\label{equ:momentum_phi}
\begin{split}
- \frac{1}{R} \frac{d}{dR} \left( R^3 \Sigma V_R \Omega \right) + \frac{1}{R} \frac{d}{dR} \left( R^3 \nu \Sigma \frac{d\Omega}{dR} \right) \\
- \frac{\left( l R \right)^2 \Omega}{2 \pi R} \frac{d\dot{M}_{\rm w}}{dR} = 0 \, ,
\end{split}
\end{equation}
\begin{equation}
\label{equ:energy}
Q_{\rm vis} = Q_{\rm adv} + Q_{\rm rad} + Q_{\rm w} \, .
\end{equation}
Eq.~(\ref{equ:continuity}) is the continuity equation, where $\Sigma$ is the surface density of the disk, $V_R$ is the radial velocity of the accretion flow, and $\dot{M}_{\rm w}$ is the total wind mass loss rate inside radius $R$. Eq.~(\ref{equ:momentum_R}) is the radial component of the momentum equation, where $\Omega$ is the angular velocity of the accretion flow, $\rho$ is the volume density, and $P$ is the total pressure, a sum of the gas, radiation, and magnetic pressures. Eq.~(\ref{equ:momentum_phi}) is the azimuthal component of the momentum equation, where $\nu$ is the kinematic viscosity coefficient, and $l$ describes the fractional amount of angular momentum carried away by the outflow. $l=1$ corresponds to the case that outflows carry away all the angular momentum that this parcel of gas occupies in the disk. Eq.~(\ref{equ:energy}) is the energy equation assuming that the disk is in thermal equilibrium, where the viscous heating rate $Q_{\rm vis}$ is balanced by the sum of advective cooling rate $Q_{\rm adv}$, radiative cooling rate $Q_{\rm rad}$, and the power carried away by the outflow $Q_{\rm w}$.

The surface density $\Sigma$ is connected with the volume density $\rho$ in the form $\Sigma = 2 \rho H$, where $H$ is the half vertical scale height of the accretion disk. The integrated mass outflow rate $\dot{M}_{\rm w}$ can be described as \citep{Knigge1999}
\begin{equation}
\label{equ:M_dot_w}
\dot{M}_{\rm w}(R) = \int_{R_{\rm in}}^R 4 \pi R^{\prime} \dot{m}_{\rm w}(R^{\prime}) dR^{\prime} \, ,
\end{equation}
where $R_{\rm in}$ is the radius of the inner edge of the accretion disk and $\dot{m}_{\rm w}$ is the mass outflow rate per unit area from the disk surface as a function of $R$. As a result of wind mass loss, we adopt the assumption that the mass accretion rate varies with radius in a power-law form \citep{Blandford1999} as
\begin{equation}
\label{equ:M_dot}
\dot{M} \equiv -2 \pi R \Sigma V_{R} = \dot{M}_0 \left( \frac{R}{R_0} \right)^p \, ,
\end{equation}
where $\dot{M}_0$ is the mass accretion rate at a specific radius $R_0$ and $p$ is a constant power law index. Substituting Eqs.~(\ref{equ:M_dot_w}, \ref{equ:M_dot}) into Eq.~(\ref{equ:continuity}), one gets
\begin{equation}
\label{equ:m_dot_w}
\dot{m}_{\rm w} = \frac{\dot{M} p}{4 \pi R^2} \, .
\end{equation}
Following \citet{Wu2022}, we assume that the power law index $p$ is proportional to the disk thickness as 
\begin{equation}
\label{equ:p}
p = \lambda(H/R) \, ,
\end{equation}
in which $\lambda$ is a constant.

The total pressure $P$ consists of the gas, radiation, and magnetic pressures:
\begin{equation}
\label{equ:pressure}
P = P_{\rm gas} + P_{\rm rad} + P_{B} \, .
\end{equation}
The gas pressure $P_{\rm gas}$ and radiation pressure $P_{\rm rad}$ are described as
\begin{equation}
P_{\rm gas} = \frac{\rho k_{\rm B}}{\mu m_{\rm p}} \left( T_{\rm i} + T_{\rm e} \right)
\end{equation}
and
\begin{equation}
P_{\rm rad} = \frac{Q_{\rm rad}}{4 c} \left( \tau + \frac{2}{\sqrt{3}} \right) \, ,
\end{equation}
where $k_{\rm B}=1.38\times10^{-16}~\rm{erg}~\rm{K}^{-1}$ is the Boltzmann constant; $\mu=0.617$ is the mean molecular weight; $T_{\rm i}$ and $T_{\rm e}$ are the ion and electron temperature, respectively, related via $T_{\rm e} = \min(T_{\rm i}, \, 6\times10^9~\rm{K})$; $\tau = (\kappa_{\rm es} + \kappa_{\rm abs})\rho H$ is the total optical depth of the disk, in which $\kappa_{\rm es}=0.34~\rm{cm}^2~\rm{g}^{-1}$ is the electron scattering opacity and $\kappa_{\rm abs}=0.27\times10^{25}\rho T_{\rm e}^{-7/2}~\rm{cm}^2~\rm{g}^{-1}$ is the absorption opacity. 

We define the sound speed associated with the gas pressure, $c_{\rm g} \equiv (P_{\rm gas} / {\rho})^{1/2}$, and that associated with the radiation pressure, $c_{\rm r} \equiv (P_{\rm rad} / {\rho})^{1/2}$. We assume that the toroidal magnetic fields will be amplified by the dynamo effect of MRI \citep{Begelman2007} until the Alfv\'{e}n velocity $V_{\rm A}$ approaches $\sqrt{V_{\rm K}c_{\rm g}}$ \citep{Pessah2005}. In this case, the magnetic pressure due to saturated magnetic fields can be described as
\begin{equation}
\label{equ:P_B}
P_{B} \equiv \rho V_{\rm A}^2 = \rho V_{\rm K} c_{\rm g} \, .
\end{equation}
This approximation has been tested with recent 3D radiative MHD simulations \citep{Huang2023}. With the inclusion of magnetic pressure, one can write the the half scale height of the disk as $H = \sqrt{c_{\rm g}^2+c_{\rm r}^2+V_{\rm A}^2}/\Omega_{\rm K}$ and the viscosity coefficient as $\nu=\alpha\sqrt{c_{\rm g}^2+c_{\rm r}^2+V_{\rm A}^2}H$, where $\alpha$ is the constant viscosity parameter.

The heating and cooling rates in the energy equation (Eq.~\ref{equ:energy}) can be expressed as \citep{Abramowicz1996, Narayan1994}
\begin{equation}
\label{equ:Q_vis}
Q_{\rm vis} = \nu \Sigma \left( R \frac{d\Omega}{dR} \right)^2 \, ,
\end{equation}
\begin{equation}
\label{equ:Q_adv}
Q_{\rm adv} = \Sigma V_R \left[ \frac{1}{\gamma - 1} \frac{d \left( c_{\rm g}^2 + c_{\rm r}^{2} + V_{\rm A}^2 \right)}{dR} - \frac{c_{\rm g}^2+ c_{\rm r}^{2} + V_{\rm A}^2}{\rho} \frac{d\rho}{dR} \right] \, ,
\end{equation}
\begin{equation}
\label{equ:Q_rad}
Q_{\rm rad} = 8 \sigma T_{\rm e}^4 \left( \frac{3\tau}{2} + \sqrt{3} + \frac{8 \sigma T_{\rm e}^4}{Q_{\rm br}^{-}} \right)^{-1} \, ,
\end{equation}
where $\gamma$ is the adiabatic index set as 1.5, $\sigma=5.67\times10^{-5}~\rm{erg}~\rm{s}^{-1}~\rm{cm}^{-2}~\rm{K}^{-4}$ is the Stefan-Boltzmann constant, $Q_{\rm br}^{-}$ is the bremsstrahlung cooling rate expressed as
\begin{equation}
Q_{\rm br}^{-} = 1.24 \times 10^{21} H \rho^2 T_{\rm e}^{1/2}~\rm{erg}~\rm{s}^{-1}~\rm{cm}^{-2} \, .
\end{equation}
Eq.~(\ref{equ:Q_rad}) can work in both optically thin and optically thick cases \citep{Narayan1995}. For the advective cooling (Eq.~\ref{equ:Q_adv}), we adopt the recipe in \citet[][see their Eq.~4]{Narayan1994}, but replace the gas pressure supported sound speed with the total pressure associated sound speed. The outflow cooling rate can be written as the wind kinetic energy per unit time,
\begin{equation}
\label{equ:Q_w}
Q_{\rm w} = 2 \eta \left(\frac{1}{2} f \dot{m}_{\rm w} V_{\rm K}^2\right) \, ,
\end{equation}
where $\eta$ is a free parameter denoting the amount of energy carried away by outflows, and $f = 1 - \left[ \Omega(3R_{\rm s}) / \Omega(R) \right] \left( 3R_{\rm s} / R \right)^{p+2}$ to correct for the zero torque at the inner boundary. The factor 2 represents the two surfaces of the accretion disk, and the $f$ parameter in it is to reconcile the fact that the viscous heating is modulated by a factor of $f$ near the inner boundary \citep{Wu2022}.

Similar to \citet{Narayan1994} and \citet{Gu2000}, we assume that the structures of the accretion disk are self-similar, i.e.,
\begin{equation}
\label{equ:self_similar}
V_R \propto R^{-\frac{1}{2}}, \, \rho \propto R^{-\frac{3}{2}}, \, \Omega \propto R^{-\frac{3}{2}}, \, c_{\rm g} \propto R^{-\frac{1}{2}}, \, c_{\rm r} \propto R^{-\frac{1}{2}} .
\end{equation}
Assuming that there is no toque at the inner stable circular orbit (ISCO), $\nu(3R_{\rm s})=0$, we substitute Eq.~(\ref{equ:continuity}) into Eq.~(\ref{equ:momentum_phi}), integrate from $3R_{\rm s}$ to $R$, and get
\begin{equation}
\label{equ:Sigma}
\nu \Sigma = \frac{\dot{M} f g^{-1}}{3 \pi} \left( 1 - \frac{l^2 p}{p + \frac{1}{2}} \right) \, ,
\end{equation}
where $g=-\frac{2}{3} \frac{d \ln{\Omega_{\rm K}}}{d \ln{R}}$. Using the self-similar assumptions in Eq.~(\ref{equ:self_similar}) and the magnetic pressure described in Eq.~(\ref{equ:P_B}), Eq.~(\ref{equ:momentum_R}) can be reduced to
\begin{equation}
\label{equ:Omega}
\frac{1}{2} V_R^2 + \frac{5}{2} (c_{\rm g}^2 + c_{\rm r}^{2}) + c_{\rm g} \Omega_{\rm K} R - c_{\rm g} R^2 \frac{d \Omega_{\rm K}}{dR} + \left( \Omega^2 - \Omega_{\rm K}^2 \right) R^2 = 0 \, .
\end{equation}
Using the self-similar assumptions and Eq.~(\ref{equ:Sigma}), Eqs.~(\ref{equ:Q_vis}, \ref{equ:Q_adv}) can be reduced to
\begin{equation}
Q_{\rm vis} = \frac{3 \dot{M} \Omega^2 f g}{4 \pi} \left( 1 - \frac{l^2 p}{p + \frac{1}{2}} \right) \, ,
\end{equation}
\begin{equation}
Q_{\rm adv} = \frac{\dot{M}}{4 \pi R^2} \left( c_{\rm g}^2 + c_{\rm r}^{2} - 5V_{\rm A}^2\right) - \frac{\dot{M} c_{\rm g}}{\pi} \frac{d\Omega_{\rm K}}{dR} \, .
\end{equation}
Substituting Eq.~(\ref{equ:m_dot_w}) into Eq.~(\ref{equ:Q_w}), one gets
\begin{equation}
Q_{\rm w} = \frac{f \eta p \dot{M} \Omega_{K}^2}{4 \pi} \, .
\end{equation}

We set $M_{\rm BH}=10M_{\sun}$, $l=1$, and $\alpha=0.1$, and leave $\dot{M}_0$, $R_0$, $\lambda$, and $\eta$ as free parameters. Now, the five equations (Eqs.~\ref{equ:energy}, \ref{equ:M_dot}, \ref{equ:pressure}, \ref{equ:Sigma}, \ref{equ:Omega}) with five variables ($\rho$, $T_{\rm i}$, $c_{\rm r}$, $\Omega$, and $V_R$) can be solved numerically.

\begin{figure*}
\centering
\includegraphics[width=0.33\textwidth]{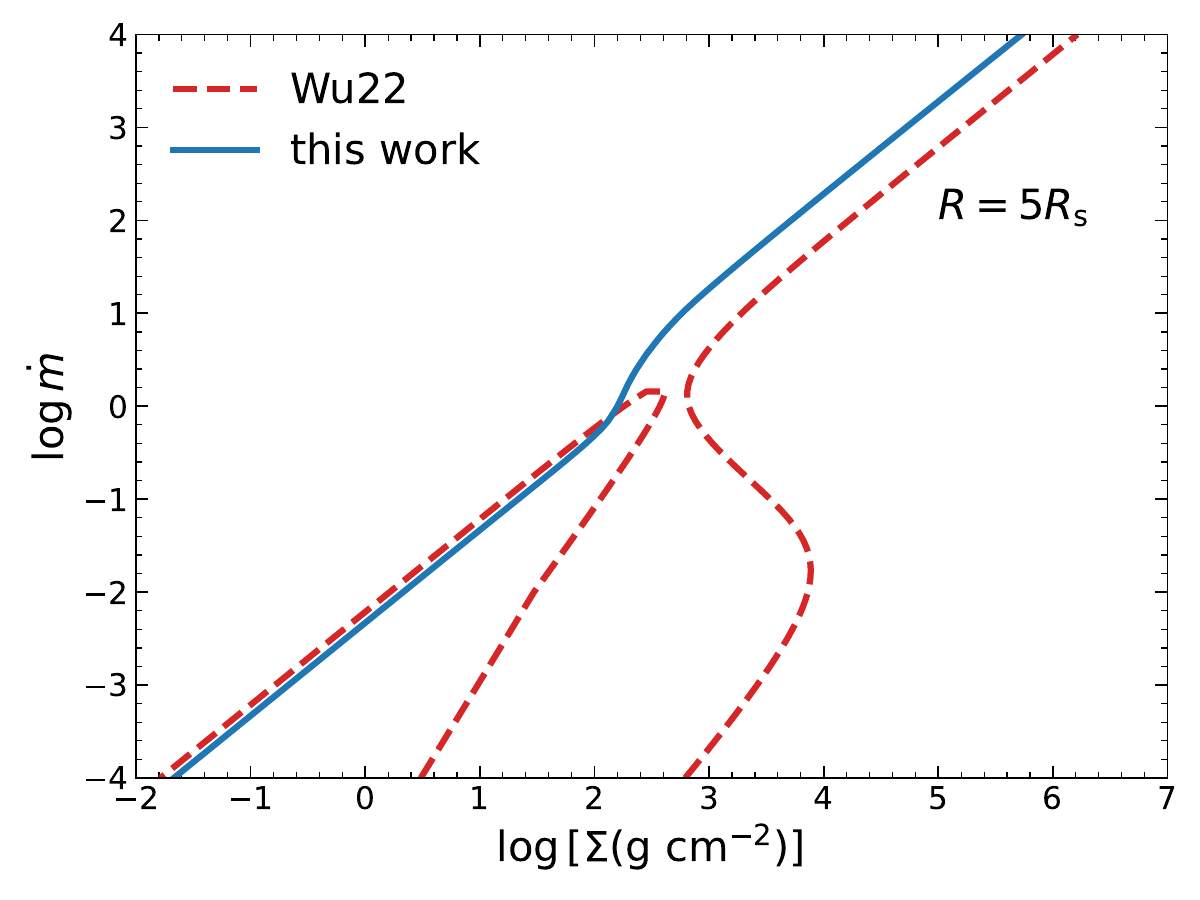}
\includegraphics[width=0.33\textwidth]{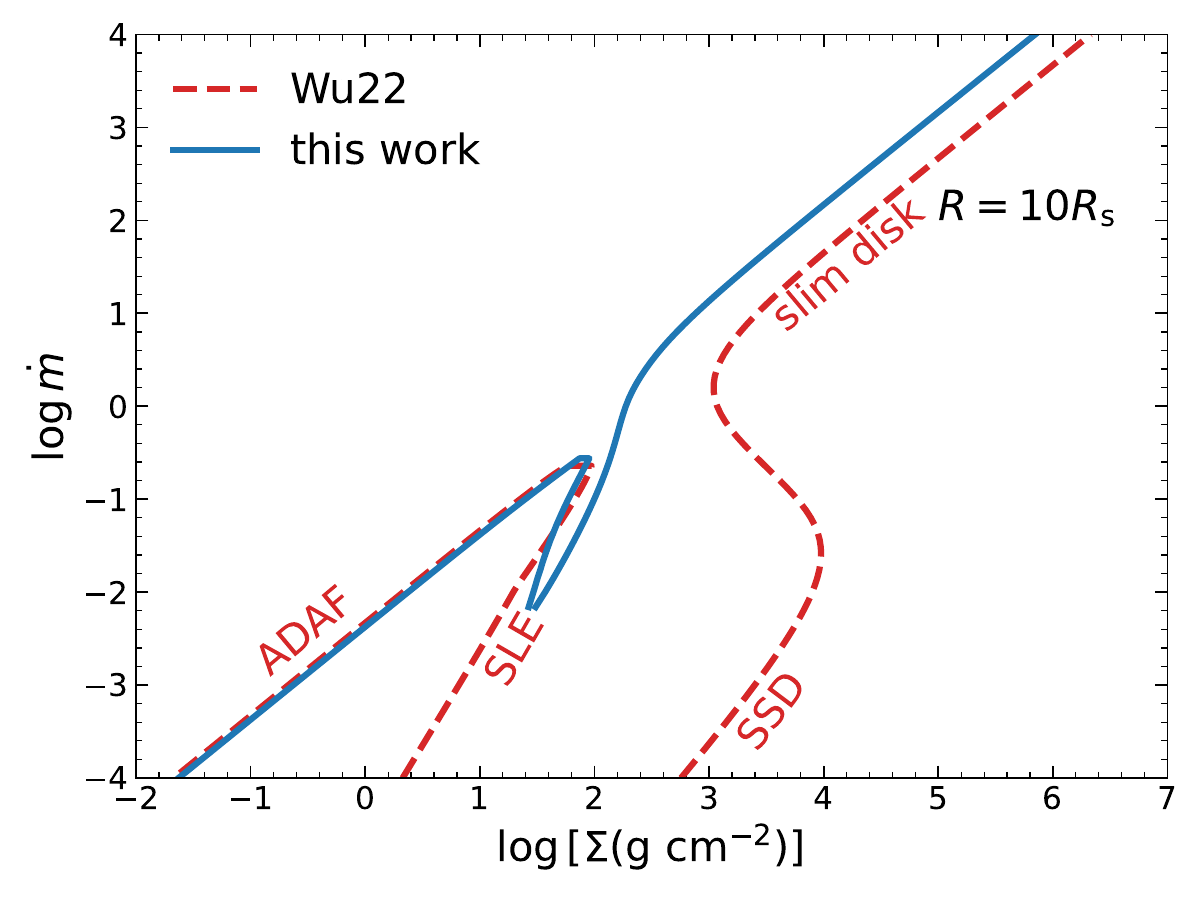}
\includegraphics[width=0.33\textwidth]{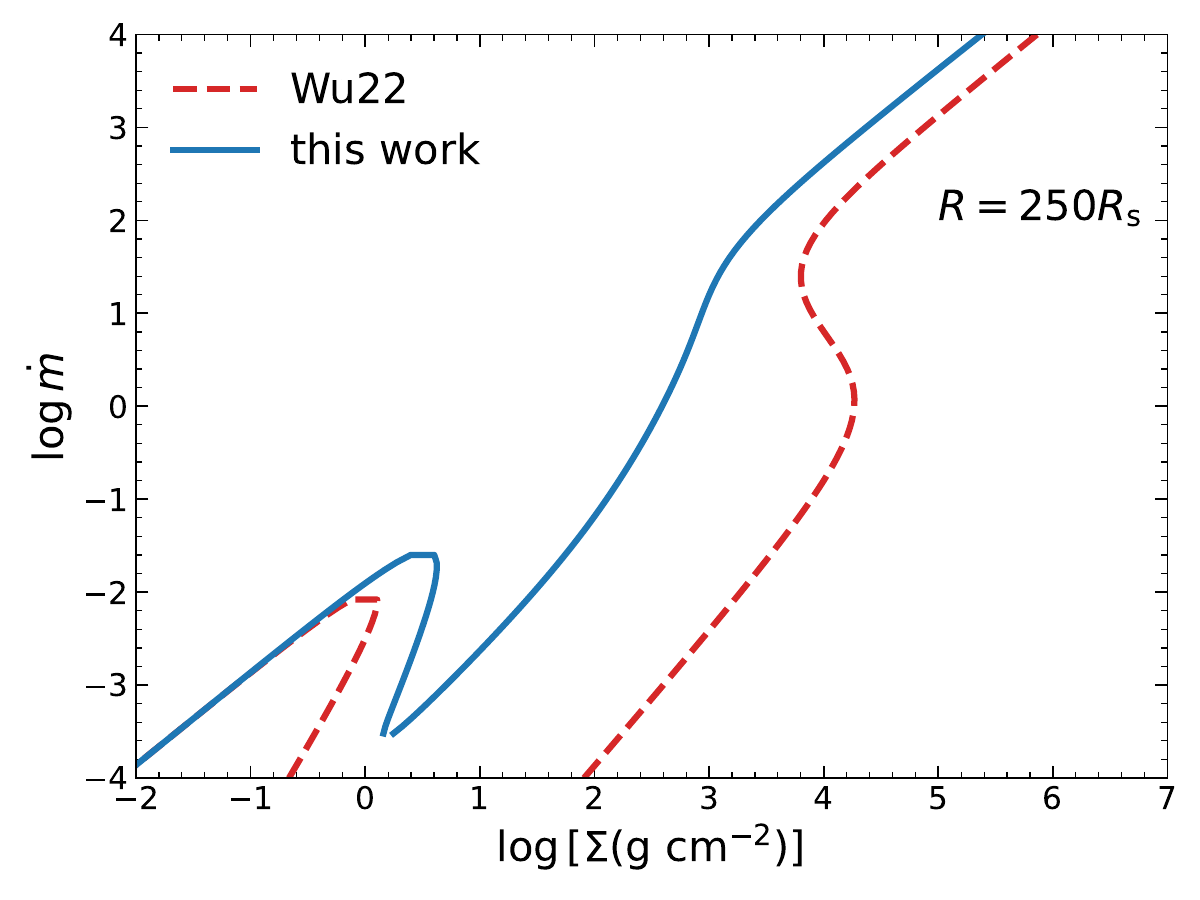}
\caption{Thermal equilibrium solutions at $R=5 R_{\rm s}$ (left), $R=10 R_{\rm s}$ (middle) and $R=250 R_{\rm s}$ (right) with $\lambda=0.5$ and $\eta=2$ on the mass accretion rate vs.\ surface density plane, in comparison with solutions from the \citet{Wu2022} model that does not consider the magnetic pressure. The four branches of solutions are marked. 
\label{fig:s_curve}}
\end{figure*}

\section{Numerical Results}
\label{sec:results}

\subsection{S-curve and effects of the magnetic pressure}
\label{sec:s-curve}

The effects of $\lambda$ and $\eta$ on the disk structure have been elaborated in \citet{Wu2022} and are similar in our model. Thus, we set $\lambda=0.5$ and $\eta=2$ as the fiducial model for the discussion of the magnetic pressure. The S-curves on the $\dot{m}$ vs.\ $\Sigma$ diagram at three radii (5, 10 and 250~$R_{\rm s}$) are shown in Figure~\ref{fig:s_curve} for our model and the \citet{Wu2022} model. Here $\dot{m}=\dot{M}/\dot{M}_{\rm crit}$, and $\dot{M}_{\rm crit}=64\pi GM_{\rm BH}/c \kappa_{\rm es}$ is defined as the critical mass accretion rate. The major difference between the two models is that a saturated magnetic pressure is included in our model. 

For the \citet{Wu2022} model, there are five branches on the S-curve seen at both radii, representing the ADAF, Shapiro-Lightman-Eardley \citep[SLE;][]{Shapiro1976}, SSD, and slim disk (see Figure~\ref{fig:s_curve}). However, our model solution at 5~$R_{\rm s}$ on the diagram reduces to a smooth (and almost straight) line that connects the ADAF branch at low accretion rates with the slim disk branch at high rates. At 10~$R_{\rm s}$ and 250~$R_{\rm s}$, the SSD branch emerges and decouples from the ADAF branch at moderate accretion rates ($10^{-3} < \dot{m} < 10^{-1}$). Interestingly, the transition from SSD to slim disk ($\dot{m} \sim 1$) no longer undergoes an unstable branch with the presence of magnetic pressure. An SLE branch appears to connect the ADAF and SSD branches, but it is found to be thermally unstable (see Section~\ref{sec:discussion}) and will not be discussed in the following. Also with magnetic pressure, the ADAF branch can reach a slightly higher accretion rate. 

\subsection{Check for self-consistency}
\label{sec:selfcons}

\begin{figure}
\centering
\includegraphics[width=0.8\columnwidth]{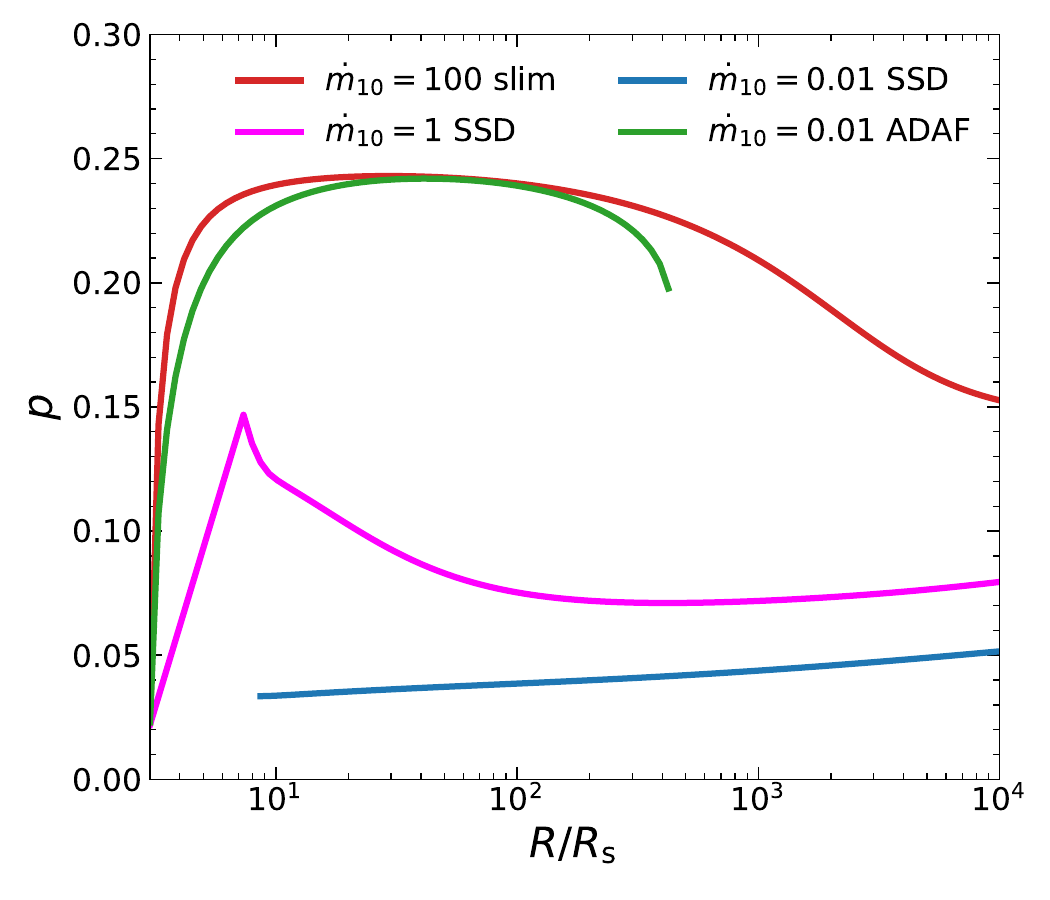}
\caption{Power-law radial dependence $p$ of the mass accretion rate as a function of radius for four solutions found at different mass accretion rates. At $\dot{m}_{10}=0.01$, the inner disk is in the ADAF branch while the outer disk is in the SSD branch. A self-consistent constant $p$ can be obtained in a wide range of radius except for the near-critical case with $\dot{m}_{10}=1$.
\label{fig:p_value}}
\end{figure}

The above solutions are found at a specific radius. The radial structure of the disk can be obtained by connecting the solutions at different radii based on the definition in Eq.~(\ref{equ:M_dot}). It is valid only if the solution satisfies the model assumption that the mass accretion rate has a power-law dependence on radius. To check that, we plot the power-law index $p$ at three typical accretion rates, $\dot{m}_{10}=0.01$, 1, and 100, in Figure~\ref{fig:p_value}, where $\dot{m}_{10}$ represents $\dot{m}$ at 10~$R_{\rm s}$. 

For $\dot{m}_{10}=0.01$, two sets of solutions can be found; the ADAF solution only exists at small radii, while the SSD solution only appears at large radii. For $\dot{m}_{10}=100$ and 1, a slim disk and SSD solution is found, respectively. As one can see, the power-law assumption, or a constant $p$, is not justified in the near-critical case with $\dot{m}_{10}=1$. For $\dot{m}_{10}=0.01$ or 100, a nearly constant $p$ is found in a wide range of radius, except for regions at small and large radii. We note that in the radial range of 5--1000~$R_{\rm s}$, the $p$ value for the ADAF or slim disk solution deviates from the median by a factor less than 15\%; for the SSD solution, the $p$ value is nearly constant at even larger radii. We will show in Section~\ref{sec:properties} that the self-similar assumptions are also valid in 5--1000~$R_{\rm s}$. Therefore, in the following of this work, we do not discuss the solution with $\dot{m}_{10}=1$, and focus on the disk properties in the radial range of 5--1000~$R_{\rm s}$. We further examined other accretion rates and found that self-consistent solutions exist at $\dot{m}_{10} < 0.5$ or $>50$.

\begin{figure*}[tb]
\centering
\includegraphics[width=0.3\textwidth]{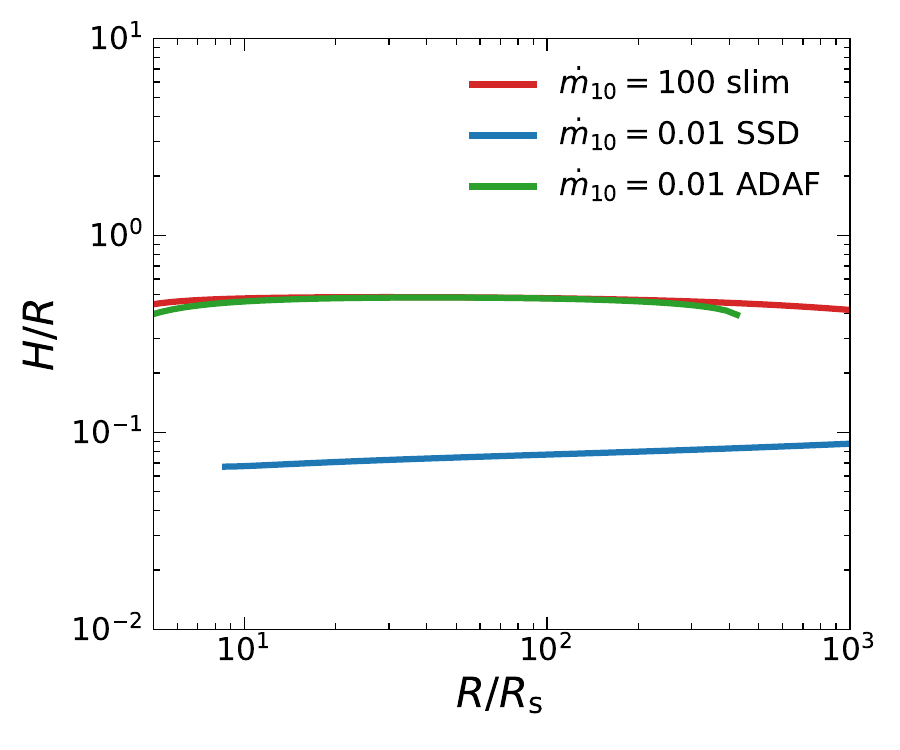}
\includegraphics[width=0.3\textwidth]{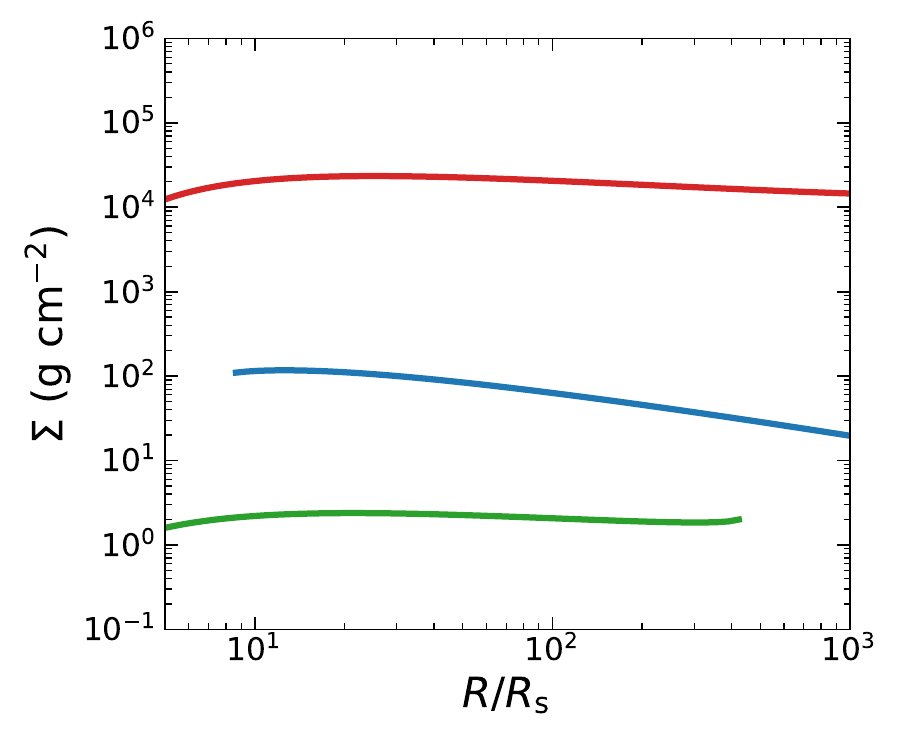}
\includegraphics[width=0.3\textwidth]{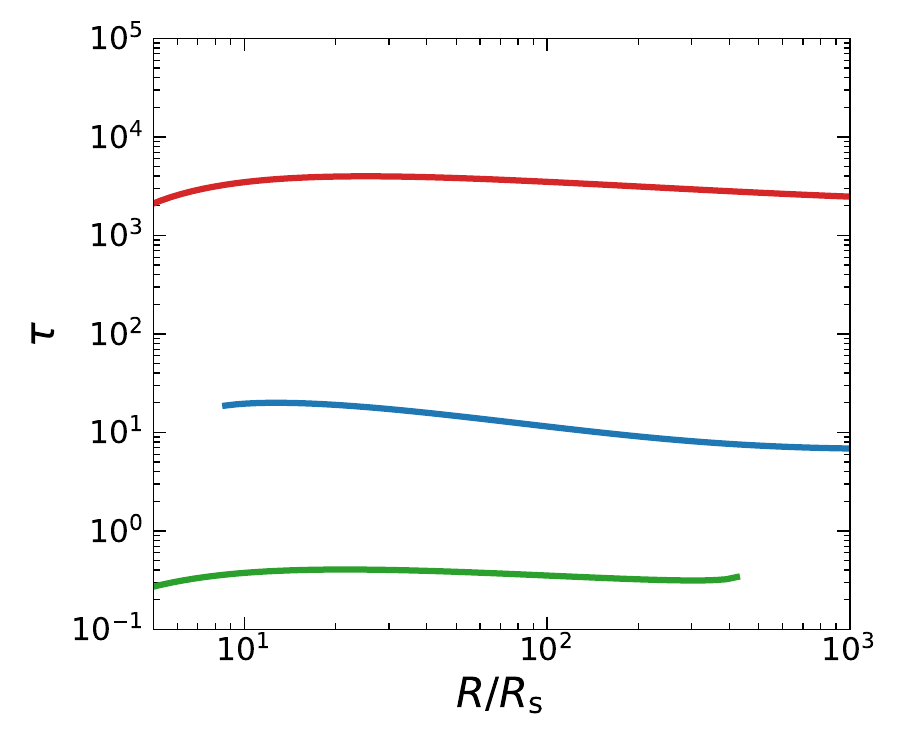}\\
\includegraphics[width=0.3\textwidth]{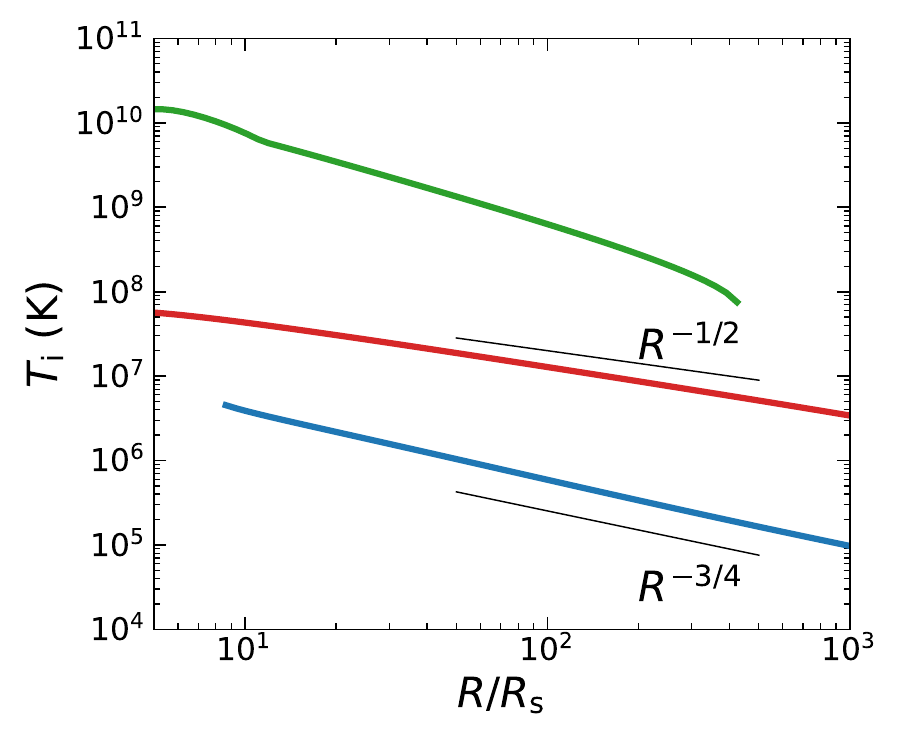}
\includegraphics[width=0.3\textwidth]{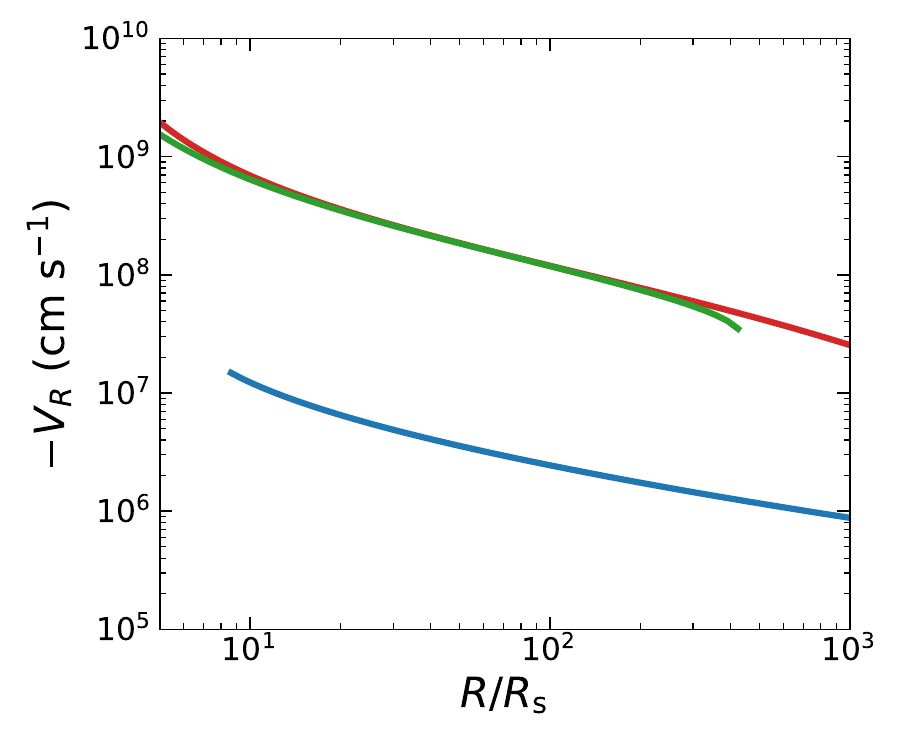}
\includegraphics[width=0.3\textwidth]{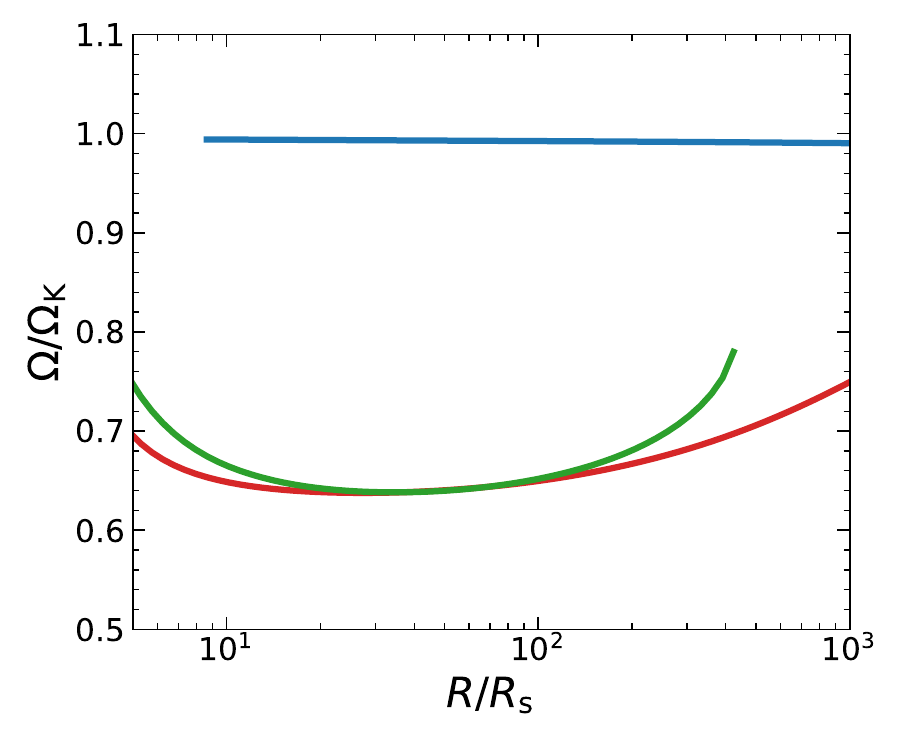}
\caption{Disk scale height ($H/R$), surface density ($\Sigma$), total optical depth ($\tau$), ion temperature ($T_{\rm i}$), radial velocity ($-V_R$), and angular velocity ($\Omega / \Omega_{\rm K}$) as a function of radius in the slim disk ($\dot{m}_{10}=100$), SSD ($\dot{m}_{10}=0.01$), and ADAF ($\dot{m}_{10}=0.01$) solutions.
\label{fig:structure}}
\end{figure*}

\begin{figure*}
\centering
\includegraphics[width=0.3\textwidth]{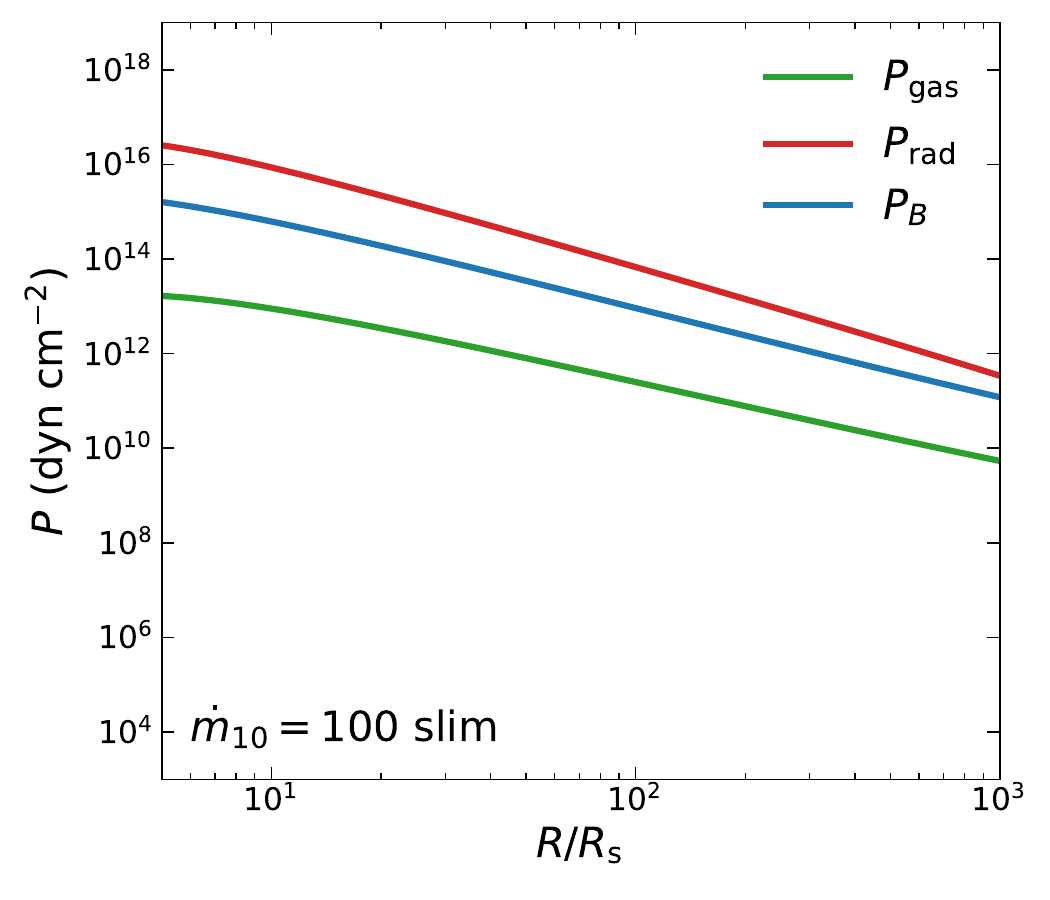}
\includegraphics[width=0.3\textwidth]{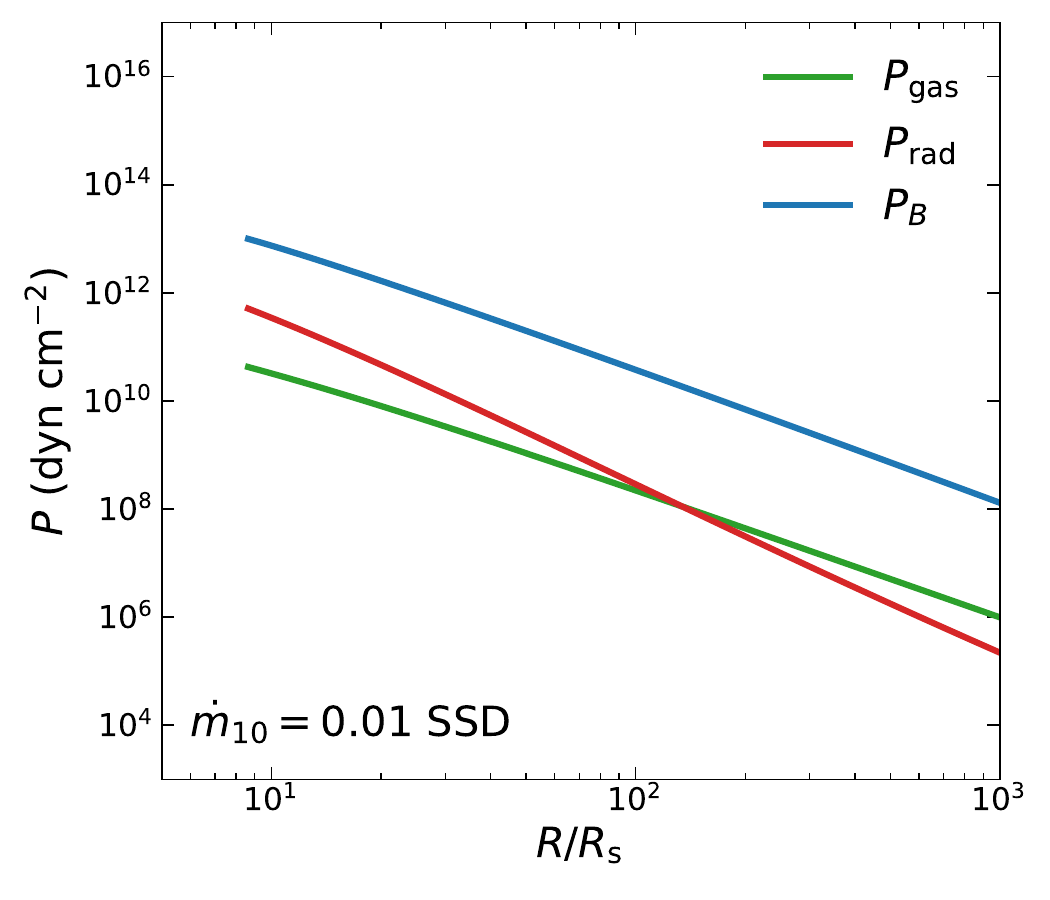}
\includegraphics[width=0.3\textwidth]{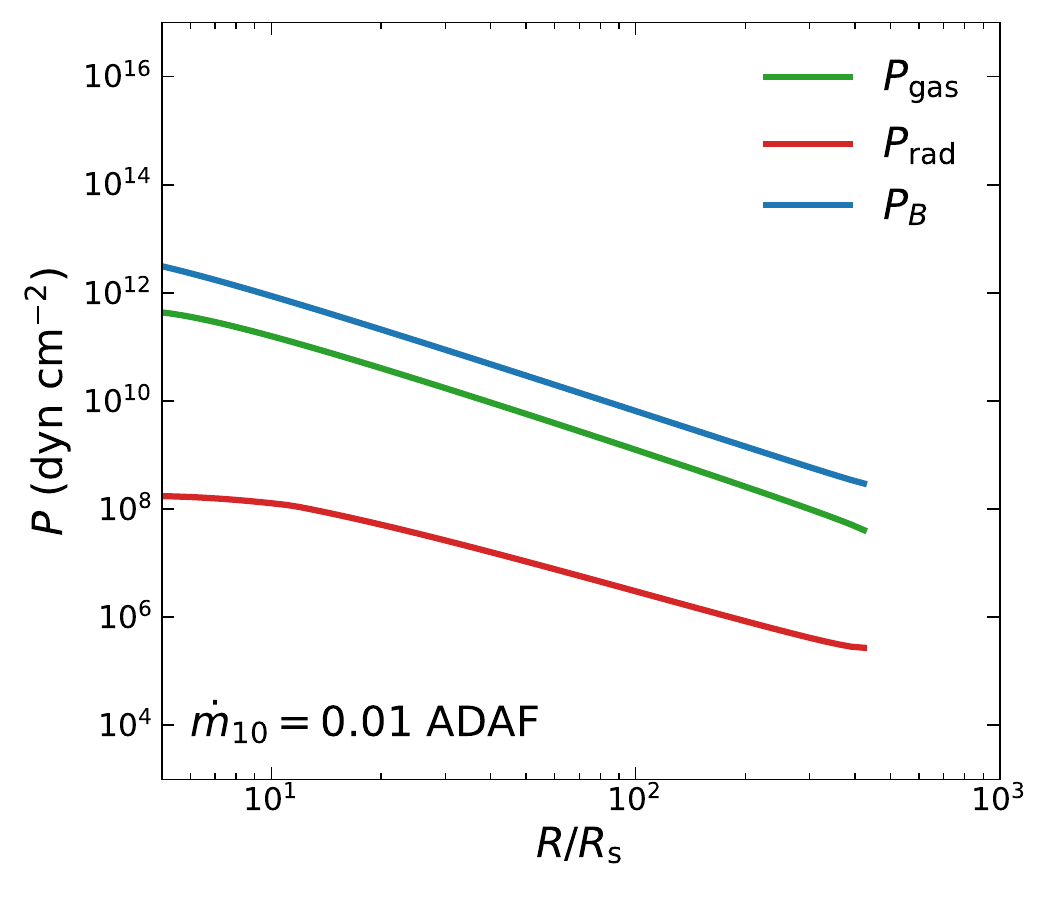}
\caption{Radial profiles of the gas, radiation, and magnetic pressures in the slim disk ($\dot{m}_{10}=100$), SSD ($\dot{m}_{10}=0.01$), and ADAF ($\dot{m}_{10}=0.01$) solutions. 
\label{fig:pressure}}
\end{figure*}

\begin{figure*}
\centering
\includegraphics[width=0.3\textwidth]{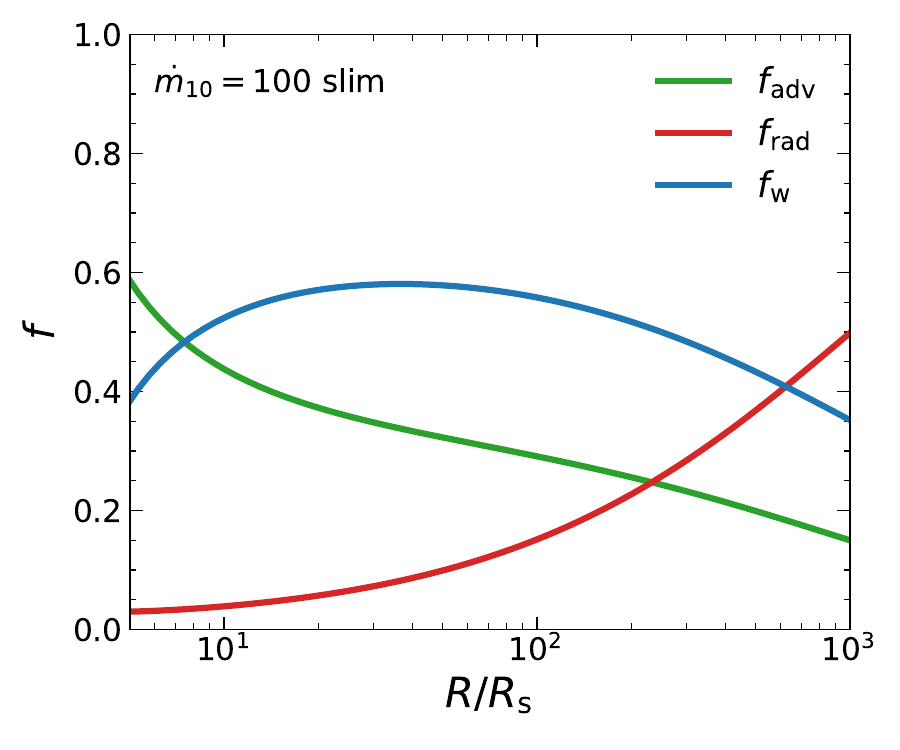}
\includegraphics[width=0.3\textwidth]{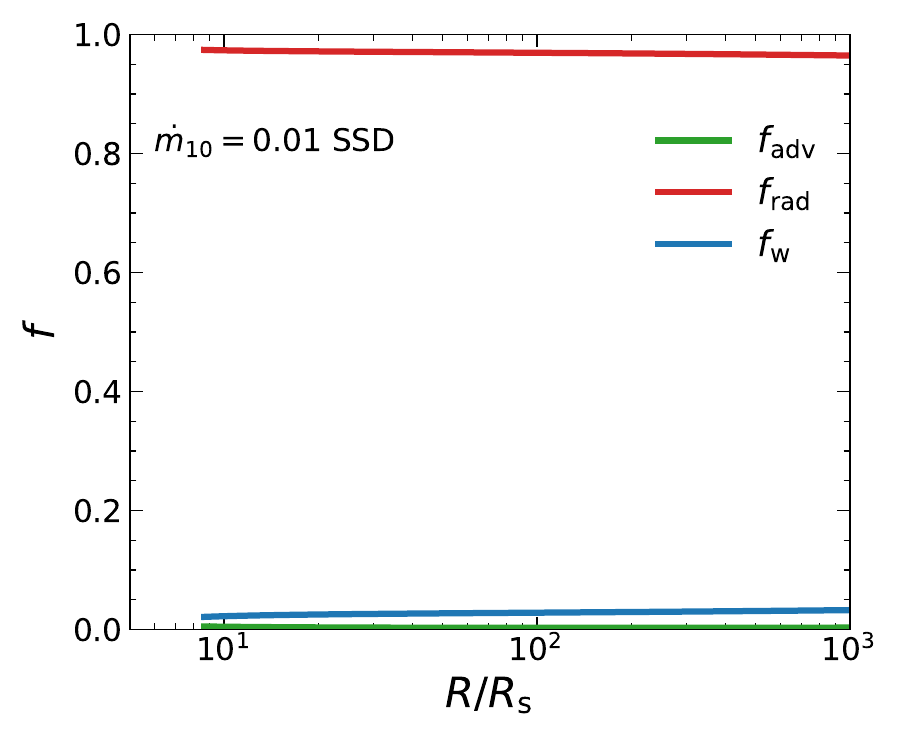}
\includegraphics[width=0.3\textwidth]{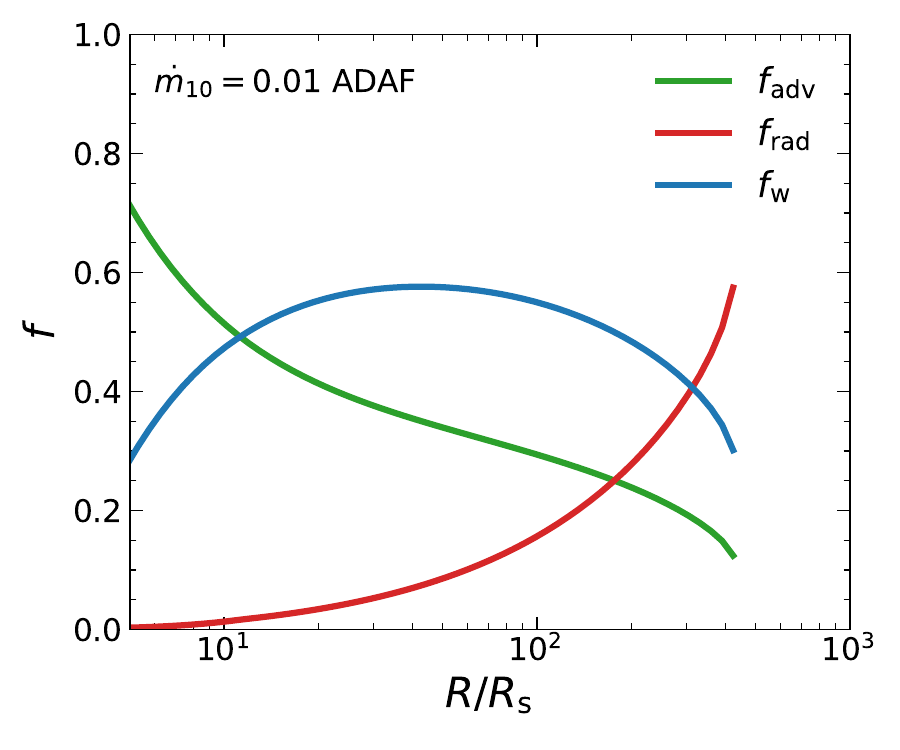}
\caption{Fractional cooling through advective ($f_{\rm adv}$), radiation ($f_{\rm rad}$), and outflow ($f_{\rm w}$) as a function of radius, in the slim disk ($\dot{m}_{10}=100$), SSD ($\dot{m}_{10}=0.01$), and ADAF ($\dot{m}_{10}=0.01$) solutions. 
\label{fig:cooling}}
\end{figure*}

\subsection{Disk properties}
\label{sec:properties}

In this section, we show the numerical results of the disk scale height ($H/R$), surface density ($\Sigma$), total optical depth ($\tau$), ion temperature ($T_{\rm i}$), radial velocity ($V_R$), and angular velocity ($\Omega / \Omega_{\rm K}$) as a function of radius in Figure~\ref{fig:structure}. For each quantity, the slim disk solution with $\dot{m}_{10}=100$ and the ADAF and SSD solutions with $\dot{m}_{10}=0.01$ are presented. 

The slim disk and ADAF solutions show similar thickness ($H/R \approx 0.5$), thicker than the SSD that has $H/R \approx 0.1$, while the slim disk has a surface density and optical depth four orders of magnitude higher than those in the ADAF solution, or two orders of magnitude higher than in SSD. As expected, the ADAF solution is optically thin and the other two are optically thick. 

The ion temperature in the ADAF solution is about two orders of magnitude higher than that in the slim disk, and the SSD is the coolest among the three. The radial temperature profile for the SSD solution is consistent with a $R^{-3/4}$ relation expected for a radiative thin disk \citep{Shakura1973}, and shows a $R^{-1/2}$ relation consistent with that expected for the slim disk model where advection becomes the dominant cooling mechanism \citep{Abramowicz1988}. 

The inflow velocities in the slim disk and ADAF are both higher than that in the SSD, due to the presence of significant advection. Similarly, they are both sub-Keplerian but the SSD shows angular velocities well consistent with the Keplerian velocity. 

We examined the self-similar assumptions (Eq.~\ref{equ:self_similar}) in the radial range of 5--1000~$R_{\rm s}$ and found no violation. The largest deviation is seen on the $\Omega$ profile at 5~$R_{\rm s}$ at a level of 9\% from the presumed $R^{-3/2}$ relation.

The radial profiles of the gas, radiation, and magnetic pressures are shown in Figure~\ref{fig:pressure} for the three solutions. The magnetic pressure is always higher than the gas pressure, which is a natural consequence of Eq.~\ref{equ:P_B}, i.e., $P_B / P_{\rm gas} \sim V_{\rm K} c_{\rm g} / c_{\rm g}^2 = V_{\rm K} / c_{\rm g} > 1$. The gas pressure is higher than the radiation pressure in the ADAF solution, and also in the SSD solution at large radii. In these two solutions, the magnetic pressure is the dominant component. In the slim disk solution, the radiation pressure becomes dominant as a result of high accretion rate.

\subsection{Cooling mechanisms}

We define the fractional cooling rate via advective, radiative, and outflow as $Q_{\rm adv}/Q_{\rm vis}$, $Q_{\rm rad}/Q_{\rm vis}$, and $Q_{\rm w}/Q_{\rm vis}$, respectively. Figure~\ref{fig:cooling} shows the variation of the three cooling channels with radius for the three solutions. The slim disk and ADAF solutions exhibit similar cooling behaviors: the advective, outflow, and radiative cooling takes over in turn from small to large radii, respectively. In particular, the outflow cooling is the dominant mechanism over most of the radii. In the SSD solution, about 95\% of the cooling goes through radiation, with the rest 5\% through outflows. 

\begin{figure}
\centering
\includegraphics[width=0.8\columnwidth]{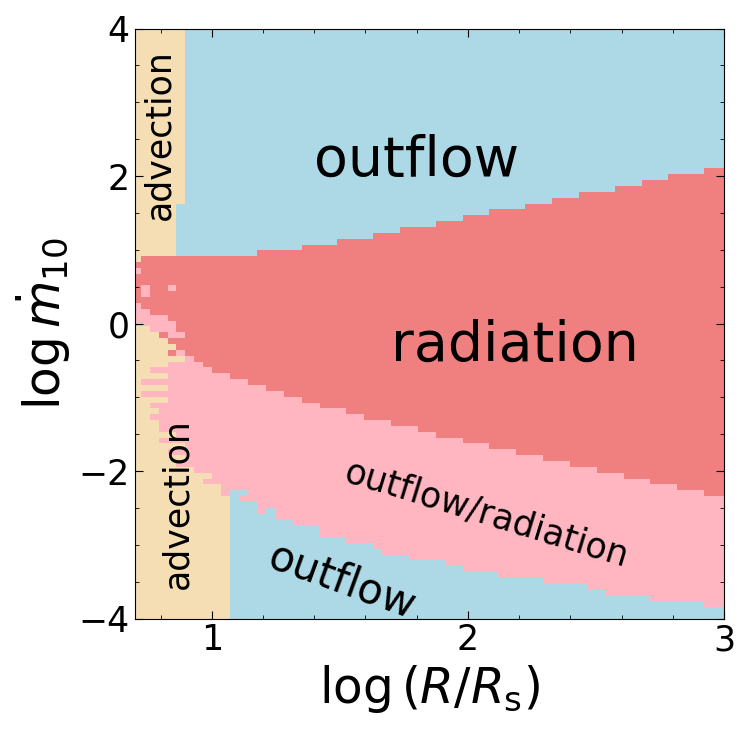}
\caption{Dominant cooling mechanism at different mass accretion rate and radius. The pink area corresponds to the region where either ADAF or SSD can exist.
\label{fig:cooling_map}}
\end{figure}

We plot the dominant cooling mechanism on the $\dot{m}$ vs.\ $R$ diagram in Figure~\ref{fig:cooling_map}. The advective cooling is the dominant component in the innermost region when the mass accretion rate is either low or high. At large radii, the outflow or radiation cooling takes over, depending on the mass accretion rate. As discussed above, the solutions around $\log \dot{m}_{10} \approx -0.3 \sim 1.7$ are not self-consistent, and the results in this region should be treated with caution.

\section{Discussion}
\label{sec:discussion}

\begin{figure*}
\centering
\includegraphics[width=0.3\textwidth]{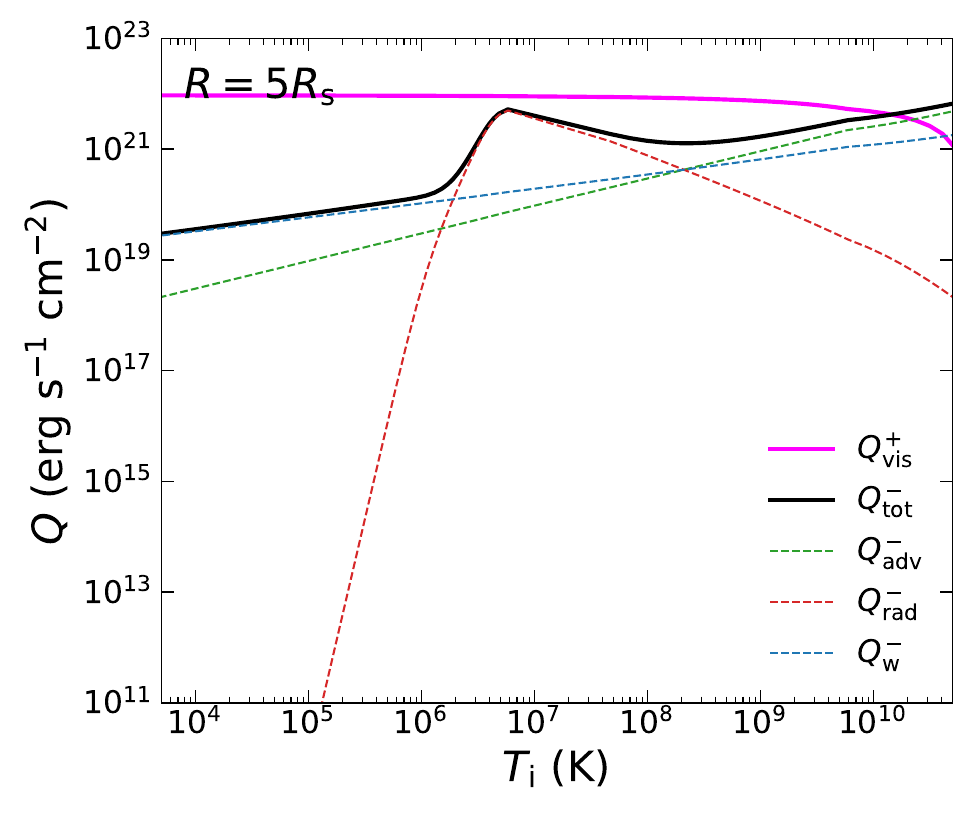}
\includegraphics[width=0.3\textwidth]{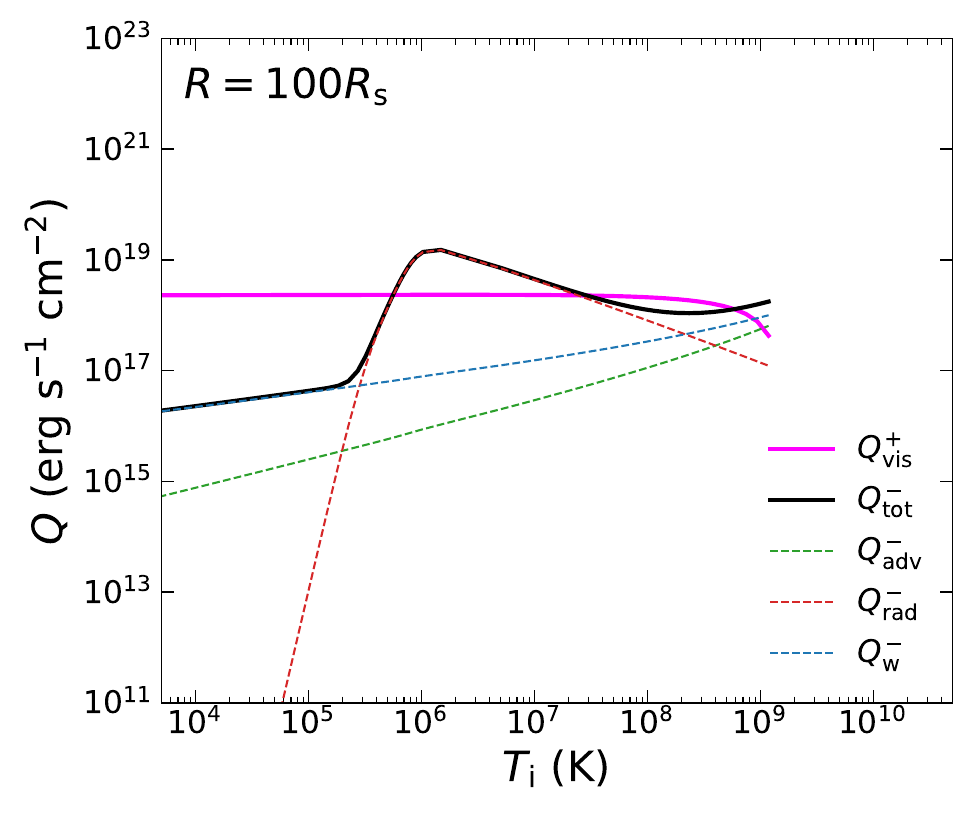}
\includegraphics[width=0.3\textwidth]{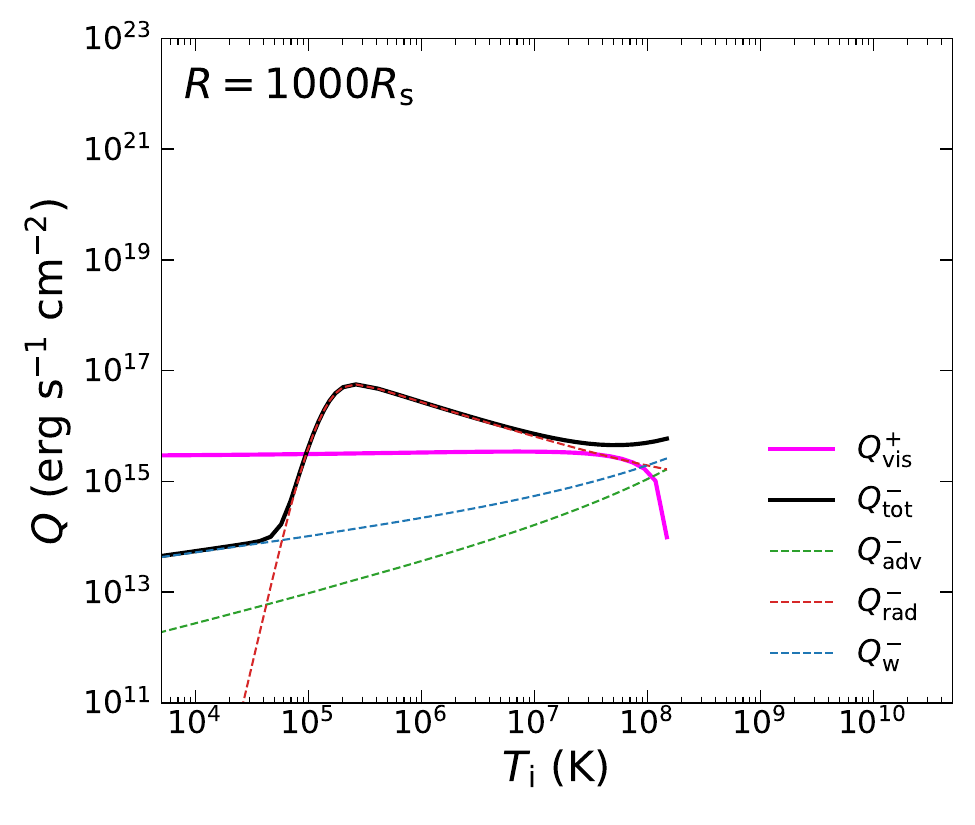}
\caption{Heating and cooling rates as a function of $T_{\rm i}$ at a mass accretion rate $\dot{m}_{10}=0.01$, calculated by relaxing the energy balance (Eq.~(\ref{equ:energy}), at three characteristic radii (5, 100, and 1000~$R_{\rm s}$). $Q_{\rm vis}^+$ is the total viscous heating rate, $Q_{\rm tot}^-$ is the total cooling rate, and $Q_{\rm adv}^-$, $Q_{\rm rad}^-$, and $Q_{\rm w}^-$ is the individual cooling component due to advection, radiation, and outflow, respectively.
\label{fig:existence}}
\end{figure*}

Inspired by the numerical simulation results \citep{Huang2023}, we construct an analytical accretion disk model including saturated magnetic pressure, as an extension to the \citet{Wu2022} model, in which outflows are already taken into account. 

\subsection{Role of magnetic pressure and outflow}

The difference between solutions of our model and those of \citet{Wu2022} reflects the role of the magnetic pressure. The main difference lies in the SSD branch. Our model predicts a much steeper radial profile for $\dot{m}$ ($p \sim 0.05$ vs.\ 0.003), indicative of enhanced outflows with the presence of  magnetic pressure. Also, the disk is much thicker with saturated magnetic pressure as a result of larger $p$ ($H/R \sim 0.1$ vs.\ 0.01). Consequently, the surface density and total optical depth drops dramatically. 

To investigate the role of outflows, we set the parameter $\lambda = 0$ and find similar disk structures with $\lambda=0.5$, indicating that the magnetic pressure plays a more important role than outflow in shaping the disk structure. Without outflow cooling, the regions where it was dominant (in the slim disk and ADAF solutions) are governed by advective cooling instead. As a result of enhanced advection, the angular velocity of the slim disk and ADAF is slightly slower, and the ADAF solution can extend to a larger radius.

\subsection{Disk truncation and an ADAF + SSD solution}

With the inclusion of magnetic pressure, there appear to have three solutions that are self-consistent with model assumptions, i.e., the slim disk solution at high accretion rates, the ADAF/SSD solutions at relatively low accretion rates. In particular, the SSD solution is only available at relatively large radii, and replaced by the ADAF solution at small radii, e.g., when $\log \dot{m}_{10} \lesssim -1$; the truncation radius grows with decreasing accretion rate (see the lower edge of the pink region in Figure~\ref{fig:cooling_map}). Observationally, a truncated SSD has been seen in black hole X-ray binaries, in particular in the hard state \citep[e.g.,][]{Churazov2001, DiSalvo2001, Homan2001, Gierlinski2008, Motta2015, Krawczynski2022, Kawamura2023}. There have been models artificially connecting a truncated SSD with an inner ADAF \citep[e.g.,][]{Esin1997, Barret2000, Done2007}. It is also suggested that the inner part of the thin disk could be evaporated and becomes a corona due to inefficient cooling, manifesting as a disk truncation \citep{Dubus2001, Liu2009, Cho2022}. Here, our solution naturally predicts the transition from a truncated SSD to an inner ADAF under a unified model.

To investigate the thermal stability of the solutions, and explain why the ADAF and SSD solutions are valid only at limited radial regions, we relax the energy equation (Eq.~\ref{equ:energy}), choose a specific ion temperature $T_{\rm i}$, and solve the remaining equations. We plot the cooling and heating rates as a function of $T_{\rm i}$ at three typical radii, namely 5, 100, and 1000~$R_{\rm s}$ in Figure~\ref{fig:existence}. We employ $\dot{m}_{10}=0.01$ such that the three radii represent the locations where only one of the ADAF/SSD solutions is available (5 and 1000~$R_{\rm s}$), or both exist (100~$R_{\rm s}$). On these plots, a solution is the intersection point between the total heating ($Q_{\rm vis}^{+}$) and cooling curve ($Q_{\rm tot}^{-}$). 

At 5~$R_{\rm s}$, the radiative cooling cannot balance the viscous heating, suggesting that the SSD solution does not exist; the only solution is to balance the viscous heating with advective plus outflow cooling near $10^{10}$~K, which is the ADAF solution. Around the intersection point, the cooling curve has a larger gradient with respect to $T_{\rm i}$ than the heating curve, suggesting that the solution is thermally stable. At 100~$R_{\rm s}$, there are two intersection points in regions where the radiative cooling dominates, standing for the SSD and SLE solutions, respectively, and another point at higher $T_{\rm i}$ with dominant outflow/advective cooling for ADAF. As one can see, the SLE solution at several $10^7$~K is thermally unstable, but both the SSD and ADAF solutions are stable. At 1000~$R_{\rm s}$, the radiative cooling can always exceed the viscous heating when $T_{\rm i} > 10^{5}$~K, causing the disk to collapse into a thin SSD disk. Again, the solution is thermally stable. 

\begin{figure}
\centering
\includegraphics[width=\columnwidth]{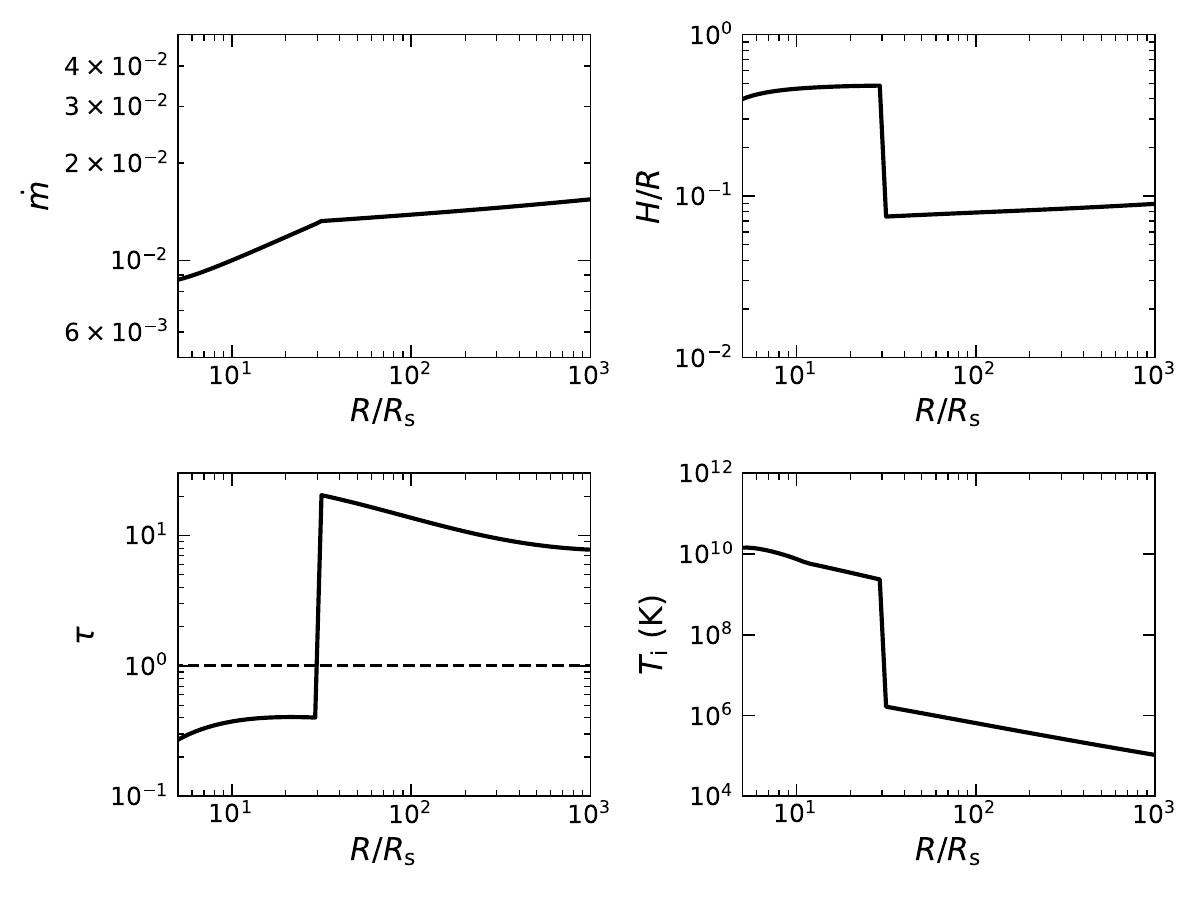}
\caption{Radial structures of a hybrid disk consisting of an outer SSD and an inner ADAF, with a transition at 30~$R_{\rm s}$ and a mass accretion rate $\dot{m}_{10}=0.01$. 
\label{fig:hybrid_model}}
\end{figure}

The coexistence of two solutions allows us to construct a hybrid disk consisting of an outer SSD and an inner ADAF. The transition radius falls into the radial range where both solutions exist, i.e., the pink region in Figure~\ref{fig:cooling_map}. For example, we set a transition radius of 30~$R_{\rm s}$ and find the two solutions along the two directions by imposing mass conservation at the transition radius. The disk structures are displayed in Figure \ref{fig:hybrid_model}. The accretion rate has a power-law form in each segment but with different power-law indices. This causes discontinuity in the disk height, optical depth, and temperature at the transition radius. In reality, the mass accretion rate around the transition radius should have a smoothed distribution, leading to a narrow but continuous transition from SSD to ADAF. Alternatively, the transition region may consist of a batch of segmented ADAFs and SSDs.

\begin{figure}
\centering
\includegraphics[width=0.8\columnwidth]{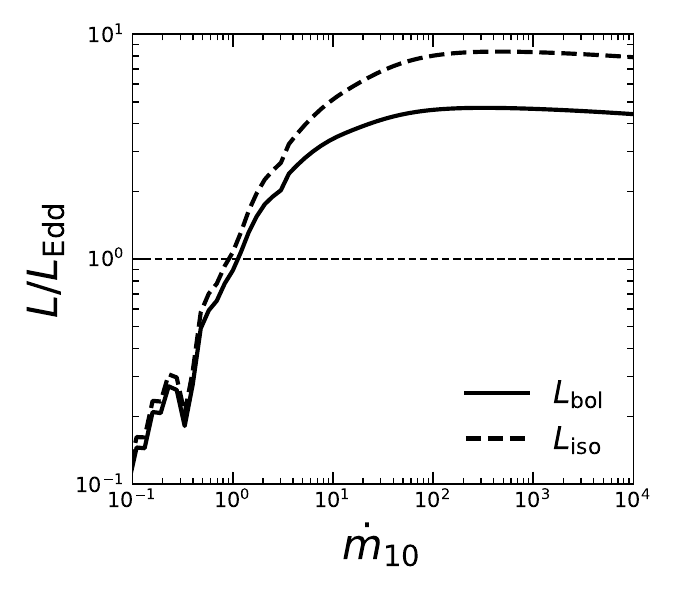}
\caption{Disk luminosity as a function of mass accretion rate. The solid curve is the bolometric luminosity and the dashed curve is the apparent luminosity assuming isotropic emission. The Eddington luminosity defined as $L_{\rm Edd}=4 \pi G M_{\rm BH} c / \kappa_{\rm es}$ is marked with a horizontal dashed line.
\label{fig:luminosity}}
\end{figure}

\subsection{Beaming effect and saturated luminosity}

When $\dot{m}_{10} > 0.1$, the whole accretion flow is expected to be optically thick and the bolometric luminosity can be integrated as
\begin{equation}
\label{equ:luminosity}
L_{\rm bol} = \int_{R_{\rm in}}^{R_{\rm out}} 2 \pi R Q_{\rm rad} dR \, ,
\end{equation}
where we set $R_{\rm in}=5R_{\rm s}$ and $R_{\rm out}=10^3R_{\rm s}$. Due to the thickening of the accretion flow, the radiation could be geometrically beamed toward the observer, and the apparent luminosity assuming isotropic emission can be expressed as
\begin{equation}
\label{equ:iso_luminosity}
L_{\rm iso} = \frac{1}{b} L_{\rm bol} \approx \frac{1}{1 - \cos{\theta}} L_{\rm bol} \, ,
\end{equation}
where $b$ is the beaming factor and $\theta$ is the half opening angle of the central funnel with $\cot{\theta}=H/R$.

The disk luminosity versus the mass accretion rate is shown in Figure~\ref{fig:luminosity}. The luminosity increases linearly with the mass accretion rate when $0.1 < \dot{m}_{10} < 5$. When the mass accretion rate further develops, the bolometric disk luminosity saturates at 4~$L_{\rm Edd}$. The luminosity of the traditional slim disk grows as a function of mass accretion rate following $(1 + \ln{\dot{m}})$ and does not have an upper limit. In our model, the apparent luminosity exceeds the Eddington limit at $\dot{m}_{10} \approx 1$, and saturates at a maximum of $\sim$8~$L_{\rm Edd}$ at $\dot{m}_{10} \ge 10^2$. This suggests that supercritical accretion onto black holes can serve as a new type of standard candle if the black hole mass is known \citep[e.g.,][]{Wang2014, King2014, Lusso2017, Negrete2018}, or one can estimate the black hole mass if supercritical accretion is evidenced.

\section*{Acknowledgements}
We thank the anonymous referee for helpful comments. HF acknowledges funding support from the National Key R\&D Project under grant 2018YFA0404502, the National Natural Science Foundation of China under grants Nos.\ 12025301, 12103027, \& 11821303, and the Tsinghua University Initiative Scientific Research Program.


\begin{thebibliography}{}
\expandafter\ifx\csname natexlab\endcsname\relax\def\natexlab#1{#1}\fi
\providecommand{\url}[1]{\href{#1}{#1}}
\providecommand{\dodoi}[1]{doi:~\href{http://doi.org/#1}{\nolinkurl{#1}}}
\providecommand{\doeprint}[1]{\href{http://ascl.net/#1}{\nolinkurl{http://ascl.net/#1}}}
\providecommand{\doarXiv}[1]{\href{https://arxiv.org/abs/#1}{\nolinkurl{https://arxiv.org/abs/#1}}}

\bibitem[{{Abramowicz} {et~al.}(1978){Abramowicz}, {Jaroszynski}, \&
  {Sikora}}]{Abramowicz1978}
{Abramowicz}, M., {Jaroszynski}, M., \& {Sikora}, M. 1978, \aap, 63, 221

\bibitem[{{Abramowicz} {et~al.}(1996){Abramowicz}, {Chen}, {Granath}, \&
  {Lasota}}]{Abramowicz1996}
{Abramowicz}, M.~A., {Chen}, X.~M., {Granath}, M., \& {Lasota}, J.~P. 1996,
  \apj, 471, 762, \dodoi{10.1086/178004}

\bibitem[{{Abramowicz} {et~al.}(1988){Abramowicz}, {Czerny}, {Lasota}, \&
  {Szuszkiewicz}}]{Abramowicz1988}
{Abramowicz}, M.~A., {Czerny}, B., {Lasota}, J.~P., \& {Szuszkiewicz}, E. 1988,
  \apj, 332, 646, \dodoi{10.1086/166683}

\bibitem[{{Balbus} \& {Hawley}(1991)}]{Balbus1991}
{Balbus}, S.~A., \& {Hawley}, J.~F. 1991, \apj, 376, 214,
  \dodoi{10.1086/170270}

\bibitem[{{Barret} {et~al.}(2000){Barret}, {Olive}, {Boirin}, {Done},
  {Skinner}, \& {Grindlay}}]{Barret2000}
{Barret}, D., {Olive}, J.~F., {Boirin}, L., {et~al.} 2000, \apj, 533, 329,
  \dodoi{10.1086/308651}

\bibitem[{{Begelman}(2012)}]{Begelman2012}
{Begelman}, M.~C. 2012, \mnras, 420, 2912,
  \dodoi{10.1111/j.1365-2966.2011.20071.x}

\bibitem[{{Begelman} \& {Pringle}(2007)}]{Begelman2007}
{Begelman}, M.~C., \& {Pringle}, J.~E. 2007, \mnras, 375, 1070,
  \dodoi{10.1111/j.1365-2966.2006.11372.x}

\bibitem[{{Blandford} \& {Begelman}(1999)}]{Blandford1999}
{Blandford}, R.~D., \& {Begelman}, M.~C. 1999, \mnras, 303, L1,
  \dodoi{10.1046/j.1365-8711.1999.02358.x}

\bibitem[{{Cao} \& {Gu}(2015)}]{Cao2015}
{Cao}, X., \& {Gu}, W.-M. 2015, \mnras, 448, 3514, \dodoi{10.1093/mnras/stv269}

\bibitem[{{Cho} \& {Narayan}(2022)}]{Cho2022}
{Cho}, H., \& {Narayan}, R. 2022, \apj, 932, 97,
  \dodoi{10.3847/1538-4357/ac6d5c}

\bibitem[{{Churazov} {et~al.}(2001){Churazov}, {Gilfanov}, \&
  {Revnivtsev}}]{Churazov2001}
{Churazov}, E., {Gilfanov}, M., \& {Revnivtsev}, M. 2001, \mnras, 321, 759,
  \dodoi{10.1046/j.1365-8711.2001.04056.x}

\bibitem[{{Di Salvo} {et~al.}(2001){Di Salvo}, {Done}, {{\.Z}ycki}, {Burderi},
  \& {Robba}}]{DiSalvo2001}
{Di Salvo}, T., {Done}, C., {{\.Z}ycki}, P.~T., {Burderi}, L., \& {Robba},
  N.~R. 2001, \apj, 547, 1024, \dodoi{10.1086/318396}

\bibitem[{{Done} {et~al.}(2007){Done}, {Gierli{\'n}ski}, \&
  {Kubota}}]{Done2007}
{Done}, C., {Gierli{\'n}ski}, M., \& {Kubota}, A. 2007, \aapr, 15, 1,
  \dodoi{10.1007/s00159-007-0006-1}

\bibitem[{{Dotan} \& {Shaviv}(2011)}]{Dotan2011}
{Dotan}, C., \& {Shaviv}, N.~J. 2011, \mnras, 413, 1623,
  \dodoi{10.1111/j.1365-2966.2011.18235.x}

\bibitem[{{Dubus} {et~al.}(2001){Dubus}, {Hameury}, \& {Lasota}}]{Dubus2001}
{Dubus}, G., {Hameury}, J.~M., \& {Lasota}, J.~P. 2001, \aap, 373, 251,
  \dodoi{10.1051/0004-6361:20010632}

\bibitem[{{Esin} {et~al.}(1997){Esin}, {McClintock}, \& {Narayan}}]{Esin1997}
{Esin}, A.~A., {McClintock}, J.~E., \& {Narayan}, R. 1997, \apj, 489, 865,
  \dodoi{10.1086/304829}

\bibitem[{{Feng} {et~al.}(2019){Feng}, {Cao}, {Gu}, \& {Ma}}]{Feng2019}
{Feng}, J., {Cao}, X., {Gu}, W.-M., \& {Ma}, R.-Y. 2019, \apj, 885, 93,
  \dodoi{10.3847/1538-4357/ab4592}

\bibitem[{{Ghoreyshi} \& {Shadmehri}(2020)}]{Ghoreyshi2020}
{Ghoreyshi}, S.~M., \& {Shadmehri}, M. 2020, \mnras, 493, 5107,
  \dodoi{10.1093/mnras/staa599}

\bibitem[{{Gierli{\'n}ski} {et~al.}(2008){Gierli{\'n}ski}, {Done}, \&
  {Page}}]{Gierlinski2008}
{Gierli{\'n}ski}, M., {Done}, C., \& {Page}, K. 2008, \mnras, 388, 753,
  \dodoi{10.1111/j.1365-2966.2008.13431.x}

\bibitem[{{Gu}(2012)}]{Gu2012}
{Gu}, W.-M. 2012, \apj, 753, 118, \dodoi{10.1088/0004-637X/753/2/118}

\bibitem[{{Gu}(2015)}]{Gu2015}
---. 2015, \apj, 799, 71, \dodoi{10.1088/0004-637X/799/1/71}

\bibitem[{{Gu} \& {Lu}(2000)}]{Gu2000}
{Gu}, W.-M., \& {Lu}, J.-F. 2000, \apjl, 540, L33, \dodoi{10.1086/312864}

\bibitem[{{Homan} {et~al.}(2001){Homan}, {Wijnands}, {van der Klis}, {Belloni},
  {van Paradijs}, {Klein-Wolt}, {Fender}, \& {M{\'e}ndez}}]{Homan2001}
{Homan}, J., {Wijnands}, R., {van der Klis}, M., {et~al.} 2001, \apjs, 132,
  377, \dodoi{10.1086/318954}

\bibitem[{{Huang} {et~al.}(2023){Huang}, {Jiang}, {Feng}, {Davis}, {Stone}, \&
  {Middleton}}]{Huang2023}
{Huang}, J., {Jiang}, Y.-F., {Feng}, H., {et~al.} 2023, \apj, 945, 57,
  \dodoi{10.3847/1538-4357/acb6fc}

\bibitem[{{Jiang} {et~al.}(2019{\natexlab{a}}){Jiang}, {Blaes}, {Stone}, \&
  {Davis}}]{Jiang2019a}
{Jiang}, Y.-F., {Blaes}, O., {Stone}, J.~M., \& {Davis}, S.~W.
  2019{\natexlab{a}}, \apj, 885, 144, \dodoi{10.3847/1538-4357/ab4a00}

\bibitem[{{Jiang} {et~al.}(2014){Jiang}, {Stone}, \& {Davis}}]{Jiang2014}
{Jiang}, Y.-F., {Stone}, J.~M., \& {Davis}, S.~W. 2014, \apj, 796, 106,
  \dodoi{10.1088/0004-637X/796/2/106}

\bibitem[{{Jiang} {et~al.}(2019{\natexlab{b}}){Jiang}, {Stone}, \&
  {Davis}}]{Jiang2019}
---. 2019{\natexlab{b}}, \apj, 880, 67, \dodoi{10.3847/1538-4357/ab29ff}

\bibitem[{{Kawamura} {et~al.}(2023){Kawamura}, {Done}, {Axelsson}, \&
  {Takahashi}}]{Kawamura2023}
{Kawamura}, T., {Done}, C., {Axelsson}, M., \& {Takahashi}, T. 2023, \mnras,
  519, 4434, \dodoi{10.1093/mnras/stad014}

\bibitem[{{King} {et~al.}(2014){King}, {Davis}, {Denney}, {Vestergaard}, \&
  {Watson}}]{King2014}
{King}, A.~L., {Davis}, T.~M., {Denney}, K.~D., {Vestergaard}, M., \& {Watson},
  D. 2014, \mnras, 441, 3454, \dodoi{10.1093/mnras/stu793}

\bibitem[{{Kitaki} {et~al.}(2017){Kitaki}, {Mineshige}, {Ohsuga}, \&
  {Kawashima}}]{Kitaki2017}
{Kitaki}, T., {Mineshige}, S., {Ohsuga}, K., \& {Kawashima}, T. 2017, \pasj,
  69, 92, \dodoi{10.1093/pasj/psx101}

\bibitem[{{Kitaki} {et~al.}(2018){Kitaki}, {Mineshige}, {Ohsuga}, \&
  {Kawashima}}]{Kitaki2018}
---. 2018, \pasj, 70, 108, \dodoi{10.1093/pasj/psy110}

\bibitem[{{Knigge}(1999)}]{Knigge1999}
{Knigge}, C. 1999, \mnras, 309, 409, \dodoi{10.1046/j.1365-8711.1999.02839.x}

\bibitem[{{Krawczynski} {et~al.}(2022){Krawczynski}, {Muleri}, {Dov{\v{c}}iak},
  {Veledina}, {Rodriguez Cavero}, {Svoboda}, {Ingram}, {Matt}, {Garcia},
  {Loktev}, {Negro}, {Poutanen}, {Kitaguchi}, {Podgorn{\'y}}, {Rankin},
  {Zhang}, {Berdyugin}, {Berdyugina}, {Bianchi}, {Blinov}, {Capitanio}, {Di
  Lalla}, {Draghis}, {Fabiani}, {Kagitani}, {Kravtsov}, {Kiehlmann},
  {Latronico}, {Lutovinov}, {Mandarakas}, {Marin}, {Marinucci}, {Miller},
  {Mizuno}, {Molkov}, {Omodei}, {Petrucci}, {Ratheesh}, {Sakanoi}, {Semena},
  {Skalidis}, {Soffitta}, {Tennant}, {Thalhammer}, {Tombesi}, {Weisskopf},
  {Wilms}, {Zhang}, {Agudo}, {Antonelli}, {Bachetti}, {Baldini}, {Baumgartner},
  {Bellazzini}, {Bongiorno}, {Bonino}, {Brez}, {Bucciantini}, {Castellano},
  {Cavazzuti}, {Ciprini}, {Costa}, {De Rosa}, {Del Monte}, {Di Gesu}, {Di
  Marco}, {Donnarumma}, {Doroshenko}, {Ehlert}, {Enoto}, {Evangelista},
  {Ferrazzoli}, {Gunji}, {Hayashida}, {Heyl}, {Iwakiri}, {Jorstad}, {Karas},
  {Kolodziejczak}, {La Monaca}, {Liodakis}, {Maldera}, {Manfreda}, {Marscher},
  {Marshall}, {Mitsuishi}, {Ng}, {O{\textquoteright}Dell}, {Oppedisano},
  {Papitto}, {Pavlov}, {Peirson}, {Perri}, {Pesce-Rollins}, {Pilia},
  {Possenti}, {Puccetti}, {Ramsey}, {Romani}, {Sgr{\`o}}, {Slane}, {Spandre},
  {Tamagawa}, {Tavecchio}, {Taverna}, {Tawara}, {Thomas}, {Trois}, {Tsygankov},
  {Turolla}, {Vink}, {Wu}, {Xie}, \& {Zane}}]{Krawczynski2022}
{Krawczynski}, H., {Muleri}, F., {Dov{\v{c}}iak}, M., {et~al.} 2022, Science,
  378, 650, \dodoi{10.1126/science.add5399}

\bibitem[{{Kumar} \& {Gu}(2018)}]{Kumar2018}
{Kumar}, R., \& {Gu}, W.-M. 2018, \apj, 860, 114,
  \dodoi{10.3847/1538-4357/aac328}

\bibitem[{{Lan{\v{c}}ov{\'a}} {et~al.}(2019){Lan{\v{c}}ov{\'a}}, {Abarca},
  {Klu{\'z}niak}, {Wielgus}, {S{\k{a}}dowski}, {Narayan}, {Schee},
  {T{\"o}r{\"o}k}, \& {Abramowicz}}]{Lancova2019}
{Lan{\v{c}}ov{\'a}}, D., {Abarca}, D., {Klu{\'z}niak}, W., {et~al.} 2019,
  \apjl, 884, L37, \dodoi{10.3847/2041-8213/ab48f5}

\bibitem[{{Lee} {et~al.}(2002){Lee}, {Reynolds}, {Remillard}, {Schulz},
  {Blackman}, \& {Fabian}}]{Lee2002}
{Lee}, J.~C., {Reynolds}, C.~S., {Remillard}, R., {et~al.} 2002, \apj, 567,
  1102, \dodoi{10.1086/338588}

\bibitem[{{Liu} \& {Taam}(2009)}]{Liu2009}
{Liu}, B.~F., \& {Taam}, R.~E. 2009, \apj, 707, 233,
  \dodoi{10.1088/0004-637X/707/1/233}

\bibitem[{{Lusso} \& {Risaliti}(2017)}]{Lusso2017}
{Lusso}, E., \& {Risaliti}, G. 2017, \aap, 602, A79,
  \dodoi{10.1051/0004-6361/201630079}

\bibitem[{{Morales Teixeira} {et~al.}(2018){Morales Teixeira}, {Avara}, \&
  {McKinney}}]{MoralesTeixeira2018}
{Morales Teixeira}, D., {Avara}, M.~J., \& {McKinney}, J.~C. 2018, \mnras, 480,
  3547, \dodoi{10.1093/mnras/sty2044}

\bibitem[{{Mosallanezhad} {et~al.}(2021){Mosallanezhad}, {Zeraatgari}, {Mei},
  \& {Bu}}]{Mosallanezhad2021}
{Mosallanezhad}, A., {Zeraatgari}, F.~Z., {Mei}, L., \& {Bu}, D.-F. 2021, \apj,
  909, 140, \dodoi{10.3847/1538-4357/abde49}

\bibitem[{{Motta} {et~al.}(2015){Motta}, {Casella}, {Henze},
  {Mu{\~n}oz-Darias}, {Sanna}, {Fender}, \& {Belloni}}]{Motta2015}
{Motta}, S.~E., {Casella}, P., {Henze}, M., {et~al.} 2015, \mnras, 447, 2059,
  \dodoi{10.1093/mnras/stu2579}

\bibitem[{{Murray} {et~al.}(1995){Murray}, {Chiang}, {Grossman}, \&
  {Voit}}]{Murray1995}
{Murray}, N., {Chiang}, J., {Grossman}, S.~A., \& {Voit}, G.~M. 1995, \apj,
  451, 498, \dodoi{10.1086/176238}

\bibitem[{{Narayan} {et~al.}(2002){Narayan}, {Quataert}, {Igumenshchev}, \&
  {Abramowicz}}]{Narayan2002}
{Narayan}, R., {Quataert}, E., {Igumenshchev}, I.~V., \& {Abramowicz}, M.~A.
  2002, \apj, 577, 295, \dodoi{10.1086/342159}

\bibitem[{{Narayan} \& {Yi}(1994)}]{Narayan1994}
{Narayan}, R., \& {Yi}, I. 1994, \apjl, 428, L13, \dodoi{10.1086/187381}

\bibitem[{{Narayan} \& {Yi}(1995)}]{Narayan1995}
---. 1995, \apj, 452, 710, \dodoi{10.1086/176343}

\bibitem[{{Negrete} {et~al.}(2018){Negrete}, {Dultzin}, {Marziani}, {Esparza},
  {Sulentic}, {del Olmo}, {Mart{\'\i}nez-Aldama}, {Garc{\'\i}a L{\'o}pez},
  {D'Onofrio}, {Bon}, \& {Bon}}]{Negrete2018}
{Negrete}, C.~A., {Dultzin}, D., {Marziani}, P., {et~al.} 2018, \aap, 620,
  A118, \dodoi{10.1051/0004-6361/201833285}

\bibitem[{{Nomura} {et~al.}(2020){Nomura}, {Ohsuga}, \& {Done}}]{Nomura2020}
{Nomura}, M., {Ohsuga}, K., \& {Done}, C. 2020, \mnras, 494, 3616,
  \dodoi{10.1093/mnras/staa948}

\bibitem[{{Nomura} {et~al.}(2016){Nomura}, {Ohsuga}, {Takahashi}, {Wada}, \&
  {Yoshida}}]{Nomura2016}
{Nomura}, M., {Ohsuga}, K., {Takahashi}, H.~R., {Wada}, K., \& {Yoshida}, T.
  2016, \pasj, 68, 16, \dodoi{10.1093/pasj/psv124}

\bibitem[{{Oda} {et~al.}(2009){Oda}, {Machida}, {Nakamura}, \&
  {Matsumoto}}]{Oda2009}
{Oda}, H., {Machida}, M., {Nakamura}, K.~E., \& {Matsumoto}, R. 2009, \apj,
  697, 16, \dodoi{10.1088/0004-637X/697/1/16}

\bibitem[{{Ohsuga} \& {Mineshige}(2011)}]{Ohsuga2011}
{Ohsuga}, K., \& {Mineshige}, S. 2011, \apj, 736, 2,
  \dodoi{10.1088/0004-637X/736/1/2}

\bibitem[{{Paczy{\'n}sky} \& {Wiita}(1980)}]{Paczynsky1980}
{Paczy{\'n}sky}, B., \& {Wiita}, P.~J. 1980, \aap, 88, 23

\bibitem[{{Pariev} {et~al.}(2003){Pariev}, {Blackman}, \&
  {Boldyrev}}]{Pariev2003}
{Pariev}, V.~I., {Blackman}, E.~G., \& {Boldyrev}, S.~A. 2003, \aap, 407, 403,
  \dodoi{10.1051/0004-6361:20030868}

\bibitem[{{Pessah} \& {Psaltis}(2005)}]{Pessah2005}
{Pessah}, M.~E., \& {Psaltis}, D. 2005, \apj, 628, 879, \dodoi{10.1086/430940}

\bibitem[{{Pinto} {et~al.}(2016){Pinto}, {Middleton}, \& {Fabian}}]{Pinto2016}
{Pinto}, C., {Middleton}, M.~J., \& {Fabian}, A.~C. 2016, \nat, 533, 64,
  \dodoi{10.1038/nature17417}

\bibitem[{{Remillard} \& {McClintock}(2006)}]{Remillard2006}
{Remillard}, R.~A., \& {McClintock}, J.~E. 2006, \araa, 44, 49,
  \dodoi{10.1146/annurev.astro.44.051905.092532}

\bibitem[{{Shakura} \& {Sunyaev}(1973)}]{Shakura1973}
{Shakura}, N.~I., \& {Sunyaev}, R.~A. 1973, \aap, 24, 337

\bibitem[{{Shapiro} {et~al.}(1976){Shapiro}, {Lightman}, \&
  {Eardley}}]{Shapiro1976}
{Shapiro}, S.~L., {Lightman}, A.~P., \& {Eardley}, D.~M. 1976, \apj, 204, 187,
  \dodoi{10.1086/154162}

\bibitem[{{S{\k{a}}dowski}(2016)}]{Sadowski2016}
{S{\k{a}}dowski}, A. 2016, \mnras, 459, 4397, \dodoi{10.1093/mnras/stw913}

\bibitem[{{S{\k{a}}dowski} \& {Narayan}(2015)}]{Sadowski2015}
{S{\k{a}}dowski}, A., \& {Narayan}, R. 2015, \mnras, 453, 3213,
  \dodoi{10.1093/mnras/stv1802}

\bibitem[{{S{\k{a}}dowski} {et~al.}(2014){S{\k{a}}dowski}, {Narayan},
  {McKinney}, \& {Tchekhovskoy}}]{Sadowski2014}
{S{\k{a}}dowski}, A., {Narayan}, R., {McKinney}, J.~C., \& {Tchekhovskoy}, A.
  2014, \mnras, 439, 503, \dodoi{10.1093/mnras/stt2479}

\bibitem[{{Stone} {et~al.}(1999){Stone}, {Pringle}, \& {Begelman}}]{stone1999}
{Stone}, J.~M., {Pringle}, J.~E., \& {Begelman}, M.~C. 1999, \mnras, 310, 1002,
  \dodoi{10.1046/j.1365-8711.1999.03024.x}

\bibitem[{{Wang} {et~al.}(2014){Wang}, {Du}, {Hu}, {Netzer}, {Bai}, {Lu},
  {Kaspi}, {Qiu}, {Li}, {Wang}, \& {SEAMBH Collaboration}}]{Wang2014}
{Wang}, J.-M., {Du}, P., {Hu}, C., {et~al.} 2014, \apj, 793, 108,
  \dodoi{10.1088/0004-637X/793/2/108}

\bibitem[{{Wang} {et~al.}(2013){Wang}, {Nowak}, {Markoff}, {Baganoff},
  {Nayakshin}, {Yuan}, {Cuadra}, {Davis}, {Dexter}, {Fabian}, {Grosso},
  {Haggard}, {Houck}, {Ji}, {Li}, {Neilsen}, {Porquet}, {Ripple}, \&
  {Shcherbakov}}]{Wang2013}
{Wang}, Q.~D., {Nowak}, M.~A., {Markoff}, S.~B., {et~al.} 2013, Science, 341,
  981, \dodoi{10.1126/science.1240755}

\bibitem[{{Wielgus} {et~al.}(2022){Wielgus}, {Lan{\v{c}}ov{\'a}}, {Straub},
  {Klu{\'z}niak}, {Narayan}, {Abarca}, {R{\'o}{\.z}a{\'n}ska}, {Vincent},
  {T{\"o}r{\"o}k}, \& {Abramowicz}}]{Wielgus2022}
{Wielgus}, M., {Lan{\v{c}}ov{\'a}}, D., {Straub}, O., {et~al.} 2022, \mnras,
  514, 780, \dodoi{10.1093/mnras/stac1317}

\bibitem[{{Wu} {et~al.}(2022){Wu}, {Gu}, \& {Sun}}]{Wu2022}
{Wu}, W.-B., {Gu}, W.-M., \& {Sun}, M. 2022, \apj, 930, 108,
  \dodoi{10.3847/1538-4357/ac6588}

\bibitem[{{Xie} \& {Yuan}(2008)}]{Xie2008}
{Xie}, F.-G., \& {Yuan}, F. 2008, \apj, 681, 499, \dodoi{10.1086/588522}

\bibitem[{{Yu} {et~al.}(2015){Yu}, {Gu}, {Liu}, {Ma}, \& {Lu}}]{Yu2015}
{Yu}, X.-F., {Gu}, W.-M., {Liu}, T., {Ma}, R.-Y., \& {Lu}, J.-F. 2015, \apj,
  801, 47, \dodoi{10.1088/0004-637X/801/1/47}

\bibitem[{{Yuan} {et~al.}(2012{\natexlab{a}}){Yuan}, {Bu}, \& {Wu}}]{Yuan2012}
{Yuan}, F., {Bu}, D., \& {Wu}, M. 2012{\natexlab{a}}, \apj, 761, 130,
  \dodoi{10.1088/0004-637X/761/2/130}

\bibitem[{{Yuan} {et~al.}(2015){Yuan}, {Gan}, {Narayan}, {Sadowski}, {Bu}, \&
  {Bai}}]{Yuan2015}
{Yuan}, F., {Gan}, Z., {Narayan}, R., {et~al.} 2015, \apj, 804, 101,
  \dodoi{10.1088/0004-637X/804/2/101}

\bibitem[{{Yuan} \& {Narayan}(2014)}]{Yuan2014}
{Yuan}, F., \& {Narayan}, R. 2014, \araa, 52, 529,
  \dodoi{10.1146/annurev-astro-082812-141003}

\bibitem[{{Yuan} {et~al.}(2012{\natexlab{b}}){Yuan}, {Wu}, \& {Bu}}]{Yuan2012a}
{Yuan}, F., {Wu}, M., \& {Bu}, D. 2012{\natexlab{b}}, \apj, 761, 129,
  \dodoi{10.1088/0004-637X/761/2/129}

\bibitem[{{Zeraatgari} {et~al.}(2021){Zeraatgari}, {Mei}, \&
  {Mosallanezhad}}]{Zeraatgari2021}
{Zeraatgari}, F.~Z., {Mei}, L., \& {Mosallanezhad}, A. 2021, \apj, 917, 19,
  \dodoi{10.3847/1538-4357/ac082d}

\bibitem[{{Zheng} {et~al.}(2011){Zheng}, {Yuan}, {Gu}, \& {Lu}}]{Zheng2011}
{Zheng}, S.-M., {Yuan}, F., {Gu}, W.-M., \& {Lu}, J.-F. 2011, \apj, 732, 52,
  \dodoi{10.1088/0004-637X/732/1/52}

\end{thebibliography}
\end{document}